\documentclass[journal]{IEEEtran}
\IEEEoverridecommandlockouts
\usepackage{cite}
\usepackage{amsmath,amssymb,amsfonts}
\usepackage{algorithmic}
\usepackage{algorithm}
\usepackage{graphicx}
\usepackage{textcomp}
\usepackage{xcolor}
\usepackage{subcaption}
\usepackage{setspace}
\usepackage{booktabs}
\usepackage{array}
\def\BibTeX{{\rm B\kern-.05em{\sc i\kern-.025em b}\kern-.08em
    T\kern-.1667em\lower.7ex\hbox{E}\kern-.125emX}}

\makeatletter

\newcommand{\Rmnum}[1]{\expandafter\@slowromancap\romannumeral #1@}
\makeatother

\makeatletter
\long\def\@makecaption#1#2{\ifx\@captype\@IEEEtablestring%
\footnotesize\begin{center}{\normalfont\footnotesize #1}\\
{\normalfont\footnotesize\scshape #2}\end{center}%
\@IEEEtablecaptionsepspace
\else
\@IEEEfigurecaptionsepspace
\setbox\@tempboxa\hbox{\normalfont\footnotesize {#1.}~~ #2}%
\ifdim \wd\@tempboxa >\hsize%
\setbox\@tempboxa\hbox{\normalfont\footnotesize {#1.}~~ }%
\parbox[t]{\hsize}{\normalfont\footnotesize \noindent\unhbox\@tempboxa#2}%
\else
\hbox to\hsize{\normalfont\footnotesize\hfil\box\@tempboxa\hfil}\fi\fi}
\makeatother

\begin{document}
\title{Energy-Efficient UAV Communications in the Presence of Wind: 3D Modeling and Trajectory Design}

\author{Xinhong~Dai,
        Bin Duo,~\IEEEmembership{Member,~IEEE},
        Xiaojun~Yuan,~\IEEEmembership{Senior~Member,~IEEE}, Marco Di Renzo,~\IEEEmembership{Fellow,~IEEE}
    \thanks{X. Dai, B. Duo and X. Yuan are with the National Key Laboratory on Wireless Communications, the University of Electronic Science and Technology of China, Chengdu 611731, China (e-mail: xhdai@std.uestc.edu.cn; duobin@cdut.edu.cn; xjyuan@uestc.edu.cn). The corresponding author is Xiaojun Yuan.}
    \thanks{M. Di Renzo is with Universit\'e Paris-Saclay, CNRS, CentraleSup\'elec, Laboratoire des Signaux et Syst\`emes, 3 Rue Joliot-Curie, 91192 Gif-sur-Yvette, France. (marco.di-renzo@universite-paris-saclay.fr).}}

\maketitle
\vspace{-1cm}
\begin{abstract}
The rapid development of unmanned aerial vehicle (UAV) technology provides flexible communication services to terrestrial nodes. Energy efficiency is crucial to the deployment of UAVs, especially rotary-wing UAVs whose propulsion power is sensitive to the wind effect. In this paper, we first derive a three-dimensional (3D) generalised propulsion energy consumption model (GPECM) for rotary-wing UAVs under the consideration of stochastic wind modeling and 3D force analysis. Based on the GPECM, we study a UAV-enabled downlink communication system, where a rotary-wing UAV flies subject to stochastic wind disturbance and provides communication services for ground users (GUs). We aim to maximize the energy efficiency (EE) of the UAV by jointly optimizing the 3D trajectory and user scheduling among the GUs based on the GPECM. We formulate the problem as stochastic optimization, which is difficult to solve due to the lack of real-time wind information. To address this issue, we propose an offline-based online adaptive (OBOA) design with two phases, namely, an offline phase and an online phase. In the offline phase, we average the wind effect on the UAV by leveraging stochastic programming (SP) based on wind statistics; then, in the online phase, we further optimize the instantaneous velocity to adapt the real-time wind. Simulation results show that the optimized trajectories of the UAV in both two phases can better adapt to the wind in changing speed and direction, and achieves a higher EE compared with the windless scheme. In particular, our proposed OBOA design can be applied in the scenario with dramatic wind changes, and makes the UAV adjust its velocity dynamically to achieve a better performance in terms of EE.
\end{abstract}

\begin{IEEEkeywords}
 Rotary-wing UAV, wind effect, generalised propulsion energy consumption model, UAV communications, energy efficiency, 3D trajectory design, stochastic programming.
\end{IEEEkeywords}

% \fi
\IEEEpeerreviewmaketitle

\section{Introduction}
Unmanned aerial vehicles (UAVs) have attracted considerable attention on providing cost-effective and diversified services in future wireless communication systems \cite{zeng2019accessing,geraci2022will,pan2021reconfigurable}. Thanks to their unique features, such as versatility, high maneuverability, and swift deployment capabilities, UAVs can easily establish line-of-sight (LoS) links with communication nodes deployed on the ground for high quality data delivery\cite{zeng2018cellular}, and can adapt to diversely complicated and time-varying environments\cite{khuwaja2018survey,mozaffari2019tutorial,di2020smart}. UAVs can be used in many applications, e.g., data collection for Internet of Things (IoTs), mobile edge computing, secure communication, and emergency communication, etc.\cite{wang2021trajectory,hayat2016survey,li2020reconfigurable,duo2020energy,9891794,li2021robust,erdelj2017help}.
\par Prior works on UAV-enabled wireless communications mostly emphasize on the placements and trajectories optimization of UAVs to improve communication performance \cite{lei2022enhancing,lyu2016placement,zeng2016throughput,boulogeorgos2022outage,yang2020performance}. However, with the diversification of missions, energy saving becomes a more and more important issue on the design of UAV-enabled communication systems. Specifically, due to the size and weight constraints, limited energy on-board significantly confines their long-term propulsion energy consumption and other prolonged operations of UAVs\cite{zeng2016wireless}. Prior research on UAV-enabled wireless communications employed simple energy consumption constraints\cite{yang2018energy,pang2021energy,su2022spectrum} that cannot effectively characterize the propulsion energy consumption of a UAV in a real environment. As such, establishing a generalised propulsion energy consumption model (GPECM) to quantify the amount of UAV energy consumption more accurately is of critical importance to the design of UAV-enabled communication systems.
\par The existing studies introduced various UAV propulsion energy consumption models from two or three-dimensional (2D or 3D) perspectives and proposed the corresponding energy-efficient designs for UAV-enabled communication systems. For rotary-wing UAVs, the authors of \cite{zeng2019energy} derived a propulsion energy consumption model in 2D space by considering UAV velocity, and optimized the trajectory to minimize the total energy consumption. In \cite{9847346}, the authors further improved the model in \cite{zeng2019energy} to include the velocity, acceleration and direction change of a UAV, and proposed a smooth and energy-efficient trajectory design. The authors of \cite{cai2022resource} provided a 3D propulsion energy consumption model, with a linear term accounting for the vertical component of UAV velocity. For fixed-wing UAVs, the authors of \cite{zeng2017energy} proposed a 2D propulsion energy consumption model taking into account UAV velocity and acceleration, where the energy efficiency (EE) was maximized for a given flight duration. The authors of \cite{xiong2022three} expanded the model in \cite{zeng2017energy} by analyzing the dynamics of the fuselage axes and derived a 3D propulsion energy consumption model that captures the impact of UAV velocity, acceleration, and pitch angle, and minimized the total energy consumption of a UAV subject to secret data collection under eavesdropping attack.
\par In general, the impact of stochastic winds cannot be neglected during UAV flight \cite{stathopoulos2018urban}. No matter flying or hovering, a UAV has to withstand the atmospheric drag caused by wind\cite{abichandani2020wind}, which causes additional propulsion energy consumption. Thus, the wind disturbance presents a real challenge to the design of UAV-enabled communication systems, especially for rotary-wing UAVs that are sensitive to the wind effect. Considering the wind effect, the authors of \cite{nachmani2007minimum} proposed an energy-saving path design of a UAV to utilize the wind. In \cite{9826413}, the authors described the wind effect on a UAV by the horizontal angle deviation between its ground velocity and the wind direction, and minimized the propulsion energy consumption subject to the given data collection constraints, by optimizing the 2D trajectories of the fixed-wing UAV.
\par As a key factor affecting UAV flight and propulsion power, the wind disturbance on UAV cannot be neglected for the energy-efficient design in UAV-enabled communication systems. However, the existing approaches either did not consider the wind effect, or did not accurately describe the impact of the wind on aerodynamics parameters (e.g.,drag and thrust) and propulsion power of UAV. Besides, real-time winds are usually random, such that traditional energy-efficient trajectory designs make limited sense in practical applications. Therefore, it is crucial to establish a GPECM to characterize the 3D movement of a UAV in wind more accurately. Based on this model, energy-efficient trajectory designs can be made possible in UAV-enabled communication systems under the stochastic wind disturbance.
\par Motivated by the above, we consider a UAV-enabled downlink communication system subject to the stochastic wind in an urban area, and propose an energy-efficient trajectory design by jointly optimizing the 3D UAV trajectory and the user scheduling. To the best of our knowledge, this is the first attempt to the 3D energy-efficient trajectory design of the UAV-enabled communication system that considers a 3D GPECM of UAV in wind. The main contributions of this paper are summarized as follows:
\begin{itemize}
    \item We propose a 3D GPECM for a rotary-wing UAV in wind based on the wind modeling and force analysis of the UAV. This model characterizes the real-time wind effect on the propulsion power and 3D movement including UAV velocity and acceleration.
    \item With the proposed GPECM, we consider a UAV-enabled urban downlink communication system. To maximize the EE of the UAV, we formulate a stochastic optimization problem by jointly optimizing 3D trajectory and user scheduling of the UAV. This problem is challenging to solve due to the difficulty in the acquirement of real-time wind information. To address this difficulty, we propose a two-phase \textit{offline-based online adaptive} (OBOA) design to achieve an energy-efficient design for the above system.
    \item In the offline phase, we assume that the statistical distribution of the wind speed and direction is known in prior. Then, we propose an offline design to maximize the EE of the UAV, by alternately optimizing the UAV's horizontal trajectory, vertical trajectory and user scheduling based on the stochastic programming (SP) technique. This offline solution is used as a reference to the optimization in the online phase.
    \item In the online phase, to make the UAV adapt better to the changeable real-time wind and reduce the impact of the wind on its propulsion energy consumption, an OBOA design is proposed to minimize the propulsion energy consumption in each time slot, by optimizing UAV instantaneous velocity slot by slot with offline trajectory taken as a reference. The OBOA design for each time slot can be implemented with relatively low complexity.
\end{itemize}
\par Extensive simulations are conducted to validate the effectiveness of our proposed design. Specifically, simulation results show that the the optimized trajectories of both offline and online phases can adapt better to the wind in changing speed and direction, and can achieve a higher EE compared to the trajectory design in a windless environment. Furthermore, with online adaptive optimization, the proposed OBOA design makes the UAV adapt better to the changeable real-time wind by dynamically adjusting its instantaneous velocity, yielding a reduced angular deviation between the UAV velocity and the real-time wind. Numerical results show that even in the scenario with dramatic wind changes, our proposed OBOA design shows a better performance in terms of EE.
\par The rest of this paper has the following structure: In Section \Rmnum{2}, the 3D GPECM in wind of the UAV is derived. In Section \Rmnum{3}, we describe the communication system model. In Section \Rmnum{4}, we formulate a problem of maximizing the EE of the UAV and propose the OBOA design. The offline and online phases are described in \Rmnum{5} and \Rmnum{6}, respectively. We discuss the simulation results in Section \Rmnum{7} and conclude this paper in Section \Rmnum{8}.
\section{3D Generalised Propulsion Energy Consumption Model in Wind}
In this section, we establish a GPECM of the UAV in wind in the 3D space. The GPECM is formulated as a function of velocity $\mathbf{v}$, acceleration $\mathbf{a}$, thrust $\mathbf{T}$ of the UAV, and wind velocity $\mathbf{v}_w$, defined as the vectors in the 3D space.
\subsection{Wind Modeling and Force Analysis}
We model the wind as a random vector, and establish the relationship between thrust $\mathbf{T}$ and other parameters by the 3D force analysis on the UAV in real-time wind. {Since the wind speed in vertical direction is commonly small \cite{calmer2018vertical}, we only consider the wind in the horizontal direction for simplicity}, and the wind vector can be expressed as $\mathbf{v}_w \triangleq\ \|\mathbf{v}_w\|\angle \beta$, where $\|\mathbf{v}_w\|$ and $\beta$ denote the speed and direction angle of the wind, respectively.
\par We model $\|\mathbf{v}_w\|$ as a random variable drawn from a Weibull distribution\cite{bowden1983weibull}, {by considering the wind speed changes with respect to (w.r.t.) altitude. Specifically, we denote $h_{\rm ref}$ and $v_{\rm ref}$ as the reference altitude and the corresponding average wind speed. Then, from \cite{yeh2008study} and \cite{tar2008some}, the wind speed at any altitude can be expressed as $\|\mathbf{v}_w\|=v_{\rm ref}\left(\frac{z}{h_{\rm ref}}\right)^p$, where $z$ is the measurement height of the wind, $p$ is an empirical exponent that depends on atmospheric conditions and $p\in [0.4,0.6]$ in the city area as shown in the following system model\cite{yeh2008study},} and $v_{\rm ref}$ follows the Weibull distribution with the probability density function (PDF)
\begin{align}
\label{weibull}
    f(v_{\rm ref}|\lambda,c) = \frac{c}{\lambda}\left(\frac{v_{\rm ref}}{\lambda}\right)^{c-1}e^{-\left(\frac{v_{\rm ref}}{\lambda}\right)^{c}},\quad \forall v_{\rm ref}>0,
\end{align}
where $\lambda$ is the scaling parameter and $c$ is the shape parameter\footnote{From \cite{bowden1983weibull}, the mean of the Weibull distribution is dominated by $\lambda$, and the variance is dominated by $c$. Therefore, to simplify the analysis, we assume that the increase of $\lambda$ represents the increase of the average wind speed, and the increase of $c$ represents the decrease of the variance of the wind speed change.}. It is worth noting that, based on our modeling, the wind effect on the UAV in the 3D space is closely related to the UAV flight height, and becomes stronger with the increasing flight height. For the wind direction, we model the direction angle $\beta$ as a random variable with a Von-Mises distribution \cite{carta2008statistical}, and the PDF of $\beta$ can be written as
\begin{align}
\label{Von-mise}
    f(\beta|\mu,\kappa) = \frac{e^{\kappa \cos{(\beta-\mu)}}}{2\pi \rm I_0(\kappa)},
\end{align}
where $\mu$ is the expectation angle, $\kappa$ is the concentration parameter and $\rm I_0(\cdot)$ denotes the zero order Bessel function\footnote{From \cite{carta2008statistical} the concentration parameter $\kappa$ denotes the variance of the Von-Mise distribution, i.e., the larger $\kappa$ means a more concentrated angle distribution with a smaller variance of the wind direction change.}. The wind at the height $z$ can be expressed as a vector:
    \begin{align}
        \label{v_w}
        \mathbf{v}_w(v_{\rm ref},\beta,z)=\left(\frac{z}{h_{\rm ref}}\right)^p(v_{\rm ref}\cos{\beta},v_{\rm ref}\sin{\beta}).
    \end{align}
    
\begin{figure}[t]
    \centering
    \includegraphics[width=9cm]{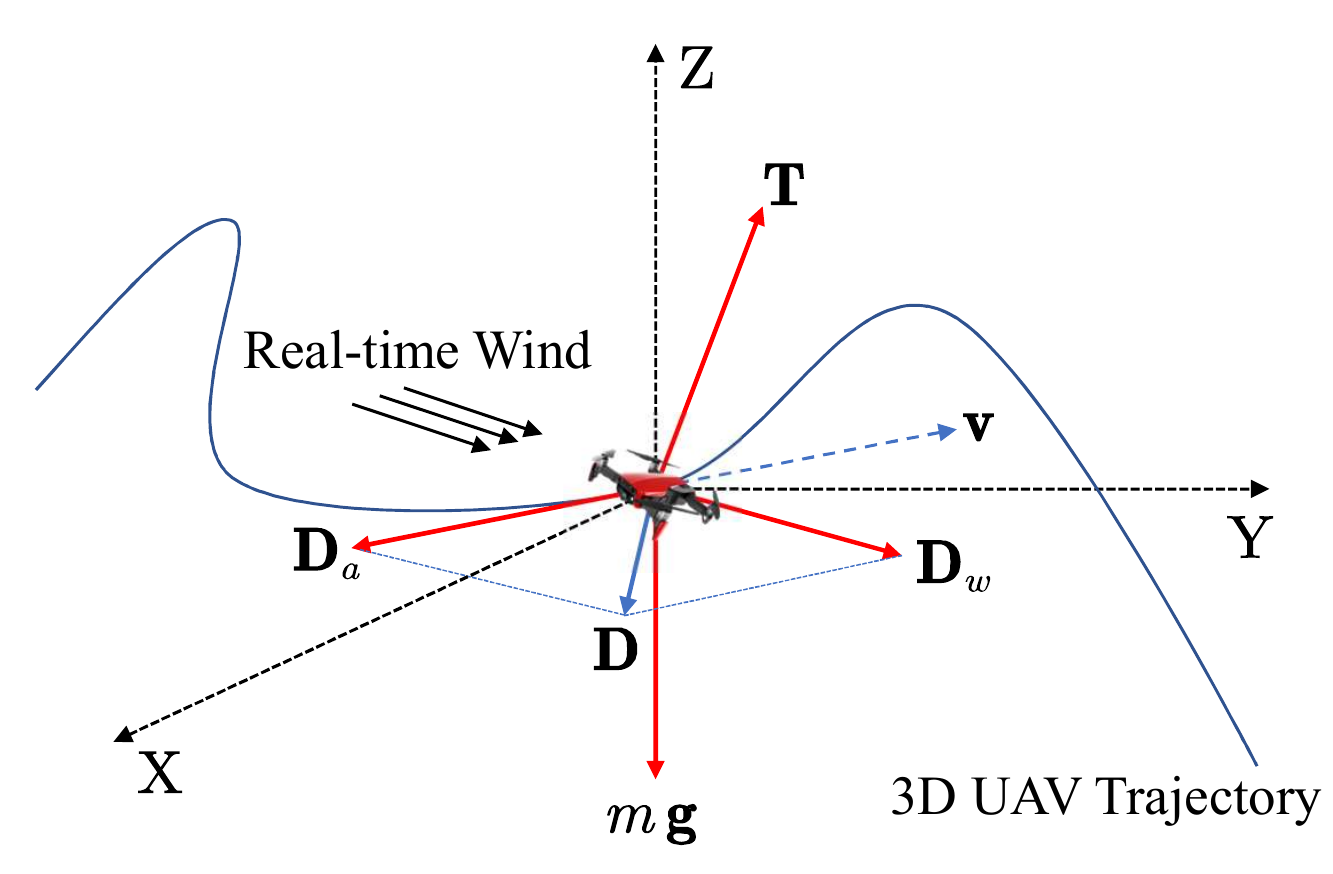}
    \caption{Instantaneous 3D force analysis on a rotary-wing UAV in real-time wind}
    \label{force analysis}
\end{figure} 
\par With the modeling of the wind, we further take a 3D force analysis of the UAV, as shown in Fig. \ref{force analysis}. Specifically, we denote the weight of the UAV and gravitational acceleration as $m$ and $\mathbf{g}$, respectively. The UAV is mainly affected by gravity $m\mathbf{g}$, thrust $\mathbf{T}$ from the rotors, air drag $\mathbf{D}_a$ due to UAV flight and the drag $\mathbf{D}_w$ caused by the wind. To simplify the analysis, we neglect the viscous effect on UAV movement of the air. Thus, $\mathbf{D}_a$ and $\mathbf{D}_w$ can be handled as two independent forces. Without loss of generality, we assume the drag caused by fluid of the UAV is isotropic. Then, the relative velocity of the UAV to the air can be denoted as $\mathbf{v}_w - \mathbf{v}$. Based on the definition of the drag in fluid \cite{bramwell2001bramwell}, the resultant drag caused by the UAV flight and wind can be expressed as
\begin{align}
    \label{D}
    \|\mathbf{D}\|= \|\mathbf{D}_a + \mathbf{D}_w\| =\frac{1}{2}\rho S_{\rm FP}\|\mathbf{v}-\mathbf{v}_w\|^2,
\end{align}
where $\rho$ and $S_{\rm FP}$ represents the air density and fuselage equivalent flat area of the UAV, respectively. Furthermore, from Newton’s second law, we have $m\mathbf{a} = \mathbf{T} + \mathbf{D} + m\mathbf{g}$. Therefore, the thrust of the UAV can be written as
\begin{align}
    \label{T}
    \|\mathbf{T}\|=\|m\mathbf{a}+\frac{1}{2}\rho S_{\rm FP}\|\mathbf{v}-\mathbf{v}_w\|(\mathbf{v}-\mathbf{v}_w)-m\mathbf{g}\|.
\end{align}
\subsection{3D GPECM Derivation}
\par With the above wind modeling and force analysis, we derive the GPECM of the UAV. Specifically, from \cite{zeng2019energy} and \cite{bramwell2001bramwell}, the propulsion power of a rotary-wing UAV can be expressed as
\begin{align}
\label{P_1}
P = q_c \rho s A v_{\rm tip}^3,
\end{align}
where the rotor solidity $s$ and disc area $A$ are both constant parameters, and $v_{\rm tip}$ denotes the tip speed of a rotor blade. From [\citenum{bramwell2001bramwell},(4.20)], the torque coefficient $q_c$ can be written as:
\begin{align}
\label{q_A_3}
q_c = \frac{\delta}{8}(1+3\hat{v}^2)+(1+c_f)\lambda_i t_c+ c_w\hat{v}\sin{\tau_c}+\frac{1}{2}d_0\hat{v}^3.
\end{align}
In addition, we have the following equations from \cite{bramwell2001bramwell}: 1) Forward speed normalized on the tip speed $\hat{v} = \frac{\|\mathbf{v}\|}{v_{\rm tip}}$; 2) thrust coefficient based on the disc axes  $t_c \approx \frac{\|\mathbf{T}\|}{\rho s A v_{\rm tip}^2}$; 3) mean induced velocity $\lambda_i = (\sqrt{v_o^4+\frac{\|\mathbf{v}\|^4}{4}}-\frac{\|\mathbf{v}\|^2}{2})^{\frac{1}{2}}$, where $v_o=\sqrt{\frac{\|\mathbf{T}\|}{2\rho A}}$ is the thrust velocity; 4) weight coefficient $c_w = \frac{\|m\mathbf{g}\|}{\rho s A v_{\rm tip}^2}$; 5) fuselage drag ratio $d_0 = \frac{S_{\rm FP}}{sA}$. According to the thrust coefficient based on the disc area $c_T$, we have $\|\mathbf{T}\|=c_T\rho A v_{\rm tip}^2$, and other parameters are detailed in [\citenum{zeng2019energy}, Table \Rmnum{1}] and \cite{bramwell2001bramwell}. With \eqref{D} and \eqref{T}, we substitute the above equations into \eqref{P_1}, yielding the UAV's 3D GPECM in wind as
\begin{align}
\label{P_2}
P = P_b + P_i +m\|\mathbf{g}\|\|\mathbf{v}\|\sin{\tau_c}+\frac{1}{2}\rho S_{\rm FP}\|\mathbf{v}-\mathbf{v}_w\|^3,
\end{align}
where $P_b = \frac{\delta}{8}(\frac{\|\mathbf{T}\|}{c_T\rho A}+3\|\mathbf{v}\|^2)\sqrt{\frac{\rho s^2A\|\mathbf{T}\|}{c_T}}$ and $P_i = (1+c_f)\|\mathbf{T}\|\left(\sqrt{\frac{\|\mathbf{T}\|^2}{(2\rho A)^2}+\frac{\|\mathbf{v}\|^4}{4}}-\frac{\|\mathbf{v}\|^2}{2}\right)^{\frac{1}{2}}$ represent the blade profile power and the induced power, respectively. The last two terms in \eqref{P_2} represent the climbing power and the power for overcoming the air drag, respectively. This GPECM characterizes the practical flight status of a rotary-wing UAV by the 3D aerodynamic analysis under the consideration of the wind, which helps to provide a more accurate energy-efficient design in the following UAV-enabled communication system. {Note that the derived GPECM \eqref{P_2} is a more general form of existing propulsion energy consumption models for rotary-wing UAVs e.g.,\cite{zeng2019energy,9847346,cai2022resource}, and it can be reduced to these models without considering the wind effect under different assumptions. The simplifications from the GPECM to the existing models in \cite{zeng2019energy,9847346,cai2022resource} are detailed in Appendix A.}
\section{System Model for UAV-Enabled Communications}
\begin{figure}[t]
    \centering
    \includegraphics[width=9cm]{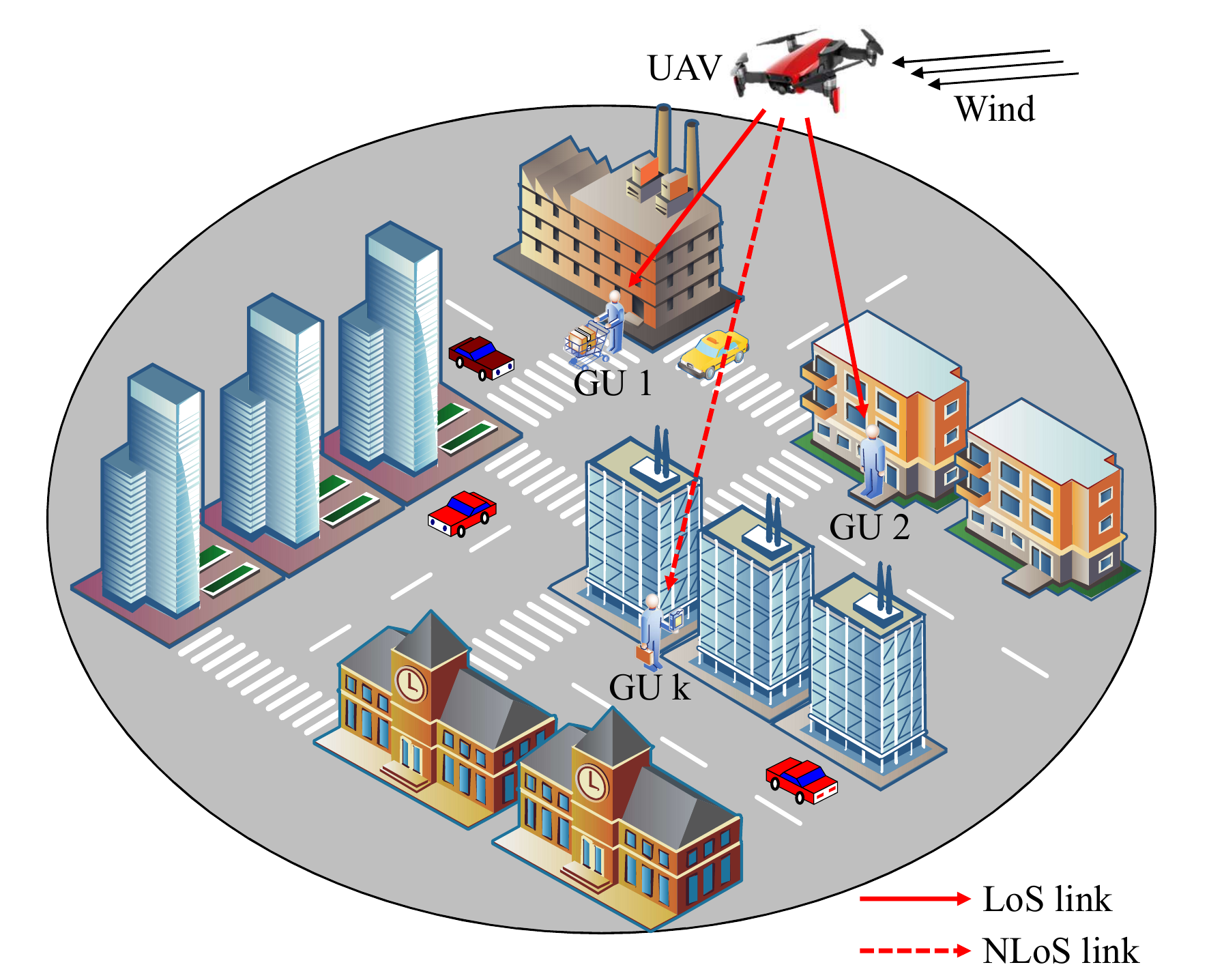}
    \caption{Downlink communication system aided by a rotary-wing UAV in wind}
    \label{system model}
\end{figure} 
As shown in Fig. \ref{system model}, we consider a general UAV-enabled downlink communication system, where a rotary-wing UAV is dispatched to fly in an urban area and provide the communication services to $K$ ground users (GUs) indexed by the set $\mathcal{K} = \{1,\dots,K\}$ in the 3D space. The UAV has to withstand strong wind with a given flight duration $T_0$. Assume that the location of each GU \emph{k} is represented by $(\mathbf{g}_k^\mathrm{T},0)$, $k \in \mathcal{K}$, where $\mathbf{g}_k = [x_k,y_k]^\mathrm{T}\in \mathbb{R}^{2\times1}$.
\subsection{Discretized UAV Mobility Model}
For simplicity, we discretize $T_0$ into $N$ equal time slots, indexed by the $\mathcal{N} = \{1,\dots,N\}$. Let $\Delta = T_0/N$ denote the each time slot, which is set sufficiently small. Denote the UAV trajectory by $\mathbf{Q}=(\mathbf{q}^\mathrm{T},z)$, where $\mathbf{q}=(x,y)^\mathrm{T}$ and $z$ represent the horizontal location and the flight altitude, respectively. As such, the UAV trajectory can be approximated by the $N$-length 3D position sequence $\{(\mathbf{q}^\mathrm{T}[n],z[n])\}_{n=1}^N$, where the choice of $N$ largely determines the complexity of 3D UAV trajectory optimization. Then, UAV velocity and acceleration can be defined as: 
\begin{align}
    \label{constraint_v_a}
    \mathbf{v}[n]\triangleq\frac{\mathbf{Q}[n+1]-\mathbf{Q}[n]}{\Delta},\quad \mathbf{a}[n]\triangleq\frac{\mathbf{Q}[n+2]+\mathbf{Q}[n]-2\mathbf{Q}[n+1]}{\Delta^2}, \quad \forall n.
\end{align}
Within each time slot $n$, we assume that the UAV location remains unchanged, and the distance between the UAV and GU $k$ is given by $d_k[n] = \sqrt{z[n]^2+\|\mathbf{q}[n]-\mathbf{g}_k\|^2}$, where $\|\cdot\|$ denotes the Euclidean norm. Moreover, $\mathbf{Q}_{\rm I}=(\mathbf{q}^\mathrm{T}[1],z[1])$ and $\mathbf{Q}_{\rm F}=(\mathbf{q}^\mathrm{T}[N],z[N])$ denote the initial and final locations of the UAV, respectively, and the UAV is launched and landed at certain locations. We assume that the horizontal and vertical speeds of the UAV are respectively limited by $V_H^{\rm max}$ and $V_V^{\rm max}$. The maximum UAV flight distance in the horizontal and vertical directions within each time slot are denoted by $D_H^{\rm max}=V_H^{\rm max}\Delta$ and $D_V^{\rm max}=V_V^{\rm max}\Delta$, respectively. Then, the UAV trajectory satisfies the following constraints:
\begin{align}
\label{constaint_v}
\|\mathbf{q}[n+1]-\mathbf{q}[n]\| \le D_H^{\rm max},\quad |z[n+1]-z[n]| \le D_V^{\rm max},\quad \forall n.
\end{align}
The flight height of the UAV is subject to the following constraints\footnote{In general, most of small drones (e.g., rotary-wing UAVs) must avoid collisions with obstacles when flying over cities, due to the highly random and changeable buildings in the city. In addition, the UAV must fly at an altitude within a legitimate range to not violate relevant aerial regulations.}
\begin{align}
\label{constaint_H}
H_{\rm min} \le z[n] \le H_{\rm max},\quad \forall n,
\end{align}
where $H_{\rm min}$ and $H_{\rm max}$ are the maximum and minimum altitudes of the UAV. In addition, based on the above discretized model, the GPECM in \eqref{P_2} can be rewritten as
\begin{align}
\label{P_3}
P[n] = P_b[n] + P_i[n] +m\|\mathbf{g}\|\|\mathbf{v}[n]\|\sin{\tau_c[n]}+\frac{1}{2}\rho S_{\rm FP}\|\mathbf{v}[n]-\mathbf{v}_w[n]\|^3,
\end{align}
where $\|\mathbf{T}[n]\|=\|m\mathbf{a}[n]+\frac{1}{2}\rho S_{\rm FP}(\mathbf{v}[n]-\mathbf{v}_w[n])-m\mathbf{g}\|$, $P_b[n] = \frac{\delta}{8}(\frac{\|\mathbf{T}[n]\|}{c_T\rho A}+3\|\mathbf{v}[n]\|^2)\sqrt{\frac{\rho s^2A\|\mathbf{T}[n]\|}{c_T}}$, and
$P_i[n] = (1+c_f)\|\mathbf{T}[n]\|\left(\sqrt{\frac{\|\mathbf{T}[n]\|^2}{(2\rho A)^2}+\frac{\|\mathbf{v}[n]\|^4}{4}}-\frac{\|\mathbf{v}[n]\|^2}{2}\right)^{\frac{1}{2}}$.
\subsection{UAV-GUs Communication Model}
 We assume that the UAV and all GUs are each equipped with a single isotropic antenna of unit gain to simplify the analysis. When the UAV communicates with the GUs in an urban area, the communication links can be occasionally blocked by buildings. Thus, the LoS link based on the free-space path loss model cannot be directly applied in our proposed system. To characterize the UAV-GUs communication channel model practically, we consider an improved probabilistic LoS channel model in \cite{you2020hybrid}, where the typical parameters of the model are obtained from a Manhattan-type city model\footnote{{Note that the channel model is divided into two categories, i.e., LoS and NLoS, and the UAV-GUS channel depends on the real environment in practice. We consider this probabilistic LoS channel model to obtain the maximum EE of the UAV from a statistical perspective in the offline phase, and consider the practical channel model in the online phase, which are detailed in the following sections.}}. Specifically, the LoS probability between the UAV and GU $k$ in time slot $n$, denoted by $P_k^L[n]$, can be rewritten as a generalized logistic function of the UAV-GU elevation angle:
\begin{align}
\label{P_L}
P_k^L[n]=A_3+\frac{A_4}{1+e^{-(A_1+A_2\theta_k[n])}},\quad \forall n, k,
\end{align}
where $A_1<0$, $A_2>0$, $A_4>0$ and $A_3$ are all constants, and $A_3+A_4 = 1$. $\theta_k[n]$ is defined as the elevation angle between the UAV and GU $k$ in time slot $n$, and 
\begin{align}
\label{theta}
\theta_k[n]=\frac{180}{\pi}\arctan\left(\frac{z[n]}{\|\mathbf{q}[n]-\mathbf{g}_k\|}\right).
\end{align}
The NLoS probability can be obtained as $P_k^N[n] = 1-P_k^L[n]$. Considering the path loss and shadowing, the channel power gains in LoS and NLoS states are expressed as 
\begin{align}
\label{channel_gain}
h_k^L[n] = \rho_0d_k^{-\alpha_L}[n],\quad h_k^N[n] = \mu_0\rho_0d_k^{-\alpha_N}[n],\quad \forall n, k,
\end{align}
where $\rho_0$ denotes the channel power gain at the unit reference distance in LoS state, $\mu_0$ is the additional signal attenuation factor due to the more complex electromagnetic propagation environment in NLoS state \cite{qian2020beamforming}. $\alpha_L$ and $\alpha_N$ denote the path loss exponents for LoS and NLoS states, respectively.

\par Assume that the UAV communicates with all GUs via the time-division multiple access (TDMA) in the same bandwidth \textit{B}, and transmits the signal with power $P_0$. We define the binary transmission scheduling variable as $\mathbf{A}=\{a_k[n],\forall n \in \mathcal{N}, k \in \mathcal{K}\}$, where the GU $k$ is served by the UAV in time slot $n$ if $a_k[n] = 1$; otherwise, $a_k[n] = 0$. Without loss of generality, we assume that the UAV can only communicate with one user in each time slot, leading to the following constraints:
\begin{align}
\label{constraint_a_0,1}
a_k[n]\in{\{0,1\}},\quad \forall n, k,
\end{align}
\begin{align}
\label{constraint_a<=1}
\sum_{k=1}^{K}a_k[n]\le1,\quad \forall n.
\end{align}
Thus, if GU $k$ is scheduled, an achievable rate of GU $k$ is given by:
\begin{align}
\label{Rate}
R_k[n] = \log_2\left(1+\frac{h_k[n]P_0}{\sigma^2\Gamma}\right),\quad \forall n,
\end{align}
where $\sigma^2$ is the power of the additive white Gaussian noise (AWGN) at the receiver, and $\Gamma$ is the signal-to-noise ratio (SNR) gap due to the practical modulation-and-coding loss. Then, the achievable rates of GU $k$ conditioned on the LoS and NLoS states are respectively given by
\begin{align}
\label{Rate_los_nlos}
R_k^L[n] = \log_2(1+\gamma_0 d_k[n]^{-\alpha_L}),\quad
R_k^N[n] = \log_2(1+\mu_0\gamma_0 d_k[n]^{-\alpha_N}),\quad \forall n,
\end{align}
where $\gamma_0 = \frac{\rho_o P_o}{\sigma^2\Gamma}$. The expected rate of GU $k$ can be expressed as
\begin{align}
\label{Rate_expected}
\mathbb{E}(R_k[n]) = P^L_k[n]R^L_k[n]+(1-P^L_k[n])R^N_k[n], \quad \forall n,
\end{align}
where $\mathbb{E}(\cdot)$ denotes the expectation.
\section{Problem Formulation}
 Considering the UAV-enabled downlink communication system where $K$ GUs are randomly assigned to the fixed locations, we aim to maximize the EE of the UAV by jointly optimizing the 3D trajectory $\mathbf{Q}$ and user scheduling $\mathbf{A}$. We assume that all GUs’ locations and the probabilistic LoS channel model are known \textit{a priori}, and the UAV can estimate the instantaneous channel state information (CSI) perfectly with individual GU in real-time along its flight\footnote{Practically, the UAV can only estimate the CSI by receiving signals from the GUs within its communication coverage. However, based on proposed system model, the UAV will not transmit information to the GUs far away from it. Thus such assumption does not compromise the practicability.}\cite{zeng2019accessing}. To ensure the fairness of the UAV communication to each GU, we introduce a relaxation variable $R_{\rm min}$, defined as the minimum threshold to the expected achievable rate of each user over $N$ time slots. Then, the original optimization problem can be formulated as: 
 \begin{subequations}\label{problem_original}
 \begin{align}
 \underset{\mathbf{Q},\mathbf{A},R_{\rm min}}{\max} \quad &\frac{R_{\rm min}}{\sum_{n=1}^{N}P[n]}\label{function_original}\\
s.t.\quad & \mathbf{Q}[1] = \mathbf{Q}_{\rm I},\quad \mathbf{Q}[N] = \mathbf{Q}_{\rm F}, \label{contraint_start_final}\\
& R_{\rm min}\le \sum_{n=1}^{N}a_k[n]\mathbb{E}(R_k[n]),\quad \forall k,\label{constraint_fairness_origin}\\
&\eqref{constaint_v},\eqref{constaint_H},\eqref{constraint_a_0,1},\eqref{constraint_a<=1},\nonumber
\end{align}
\end{subequations}
since $P[n]$ and \eqref{constraint_fairness_origin} are non-convex w.r.t. $\mathbf{Q}$, and \eqref{constraint_a_0,1} gives integer constraints, problem \eqref{problem_original} is a mixed-integer non-convex fractional program, which is challenging to solve in general. Additionally, problem \eqref{problem_original} is difficult to handle due to the lack of real-time wind information in $P[n]$, i.e., speeds and directions in all time slots prior to flight. As such, we propose an OBOA design with offline and online phases, shown in Fig. \ref{OBOA}\footnote{{From \cite{zeng2019accessing}, the proposed algorithm in the offline phase can be executed on a ground control center, which has enough computing capability and can exchange information and control signal to the UAV. Then, the proposed algorithm in the online phase can be executed in a UAV-mounted computer.}}, to reduce the real-time wind impact on UAV flight and obtain an energy-efficient solution to problem \eqref{problem_original}. Specifically, the general descriptions of the two phases are as follows:
\begin{itemize}
    \item \textit{Offline phase}: To solve problem \eqref{problem_original}, before the flight, we assume that the statistical distributions of both the wind speed and direction are known\footnote{{Before the flight, the UAV can be launched to hover at a fixed altitude to collect wind information based on the UAV-mounted electronic/mechanical anemometer and wind direction indicator or other high precision sensors. Then, the parameters of the proposed distributions can be estimated by existing estimation algorithms e.g., maximum likelihood estimation. Similarly, the UAV can also collect real-time wind information based on the mounted sensors in the online phase. The proposed design can be extended to address other distributions of the wind.}}. With given wind statistics, we propose an offline design to average the wind effect on the UAV, by jointly optimizing the 3D trajectory $\mathbf{Q}$ and user scheduling $\mathbf{A}$. The proposed offline design based on SP yields a maximum EE of the UAV from a statistically favorable perspective. Note that the offline design is obtained based on the wind statistics instead of the real-time wind information. Therefore, the offline design leads to a suboptimal solution to problem \eqref{problem_original} which is used as a reference for the further online adaptive optimization.
\begin{figure}[t]
    \centering
    \includegraphics[width=14cm]{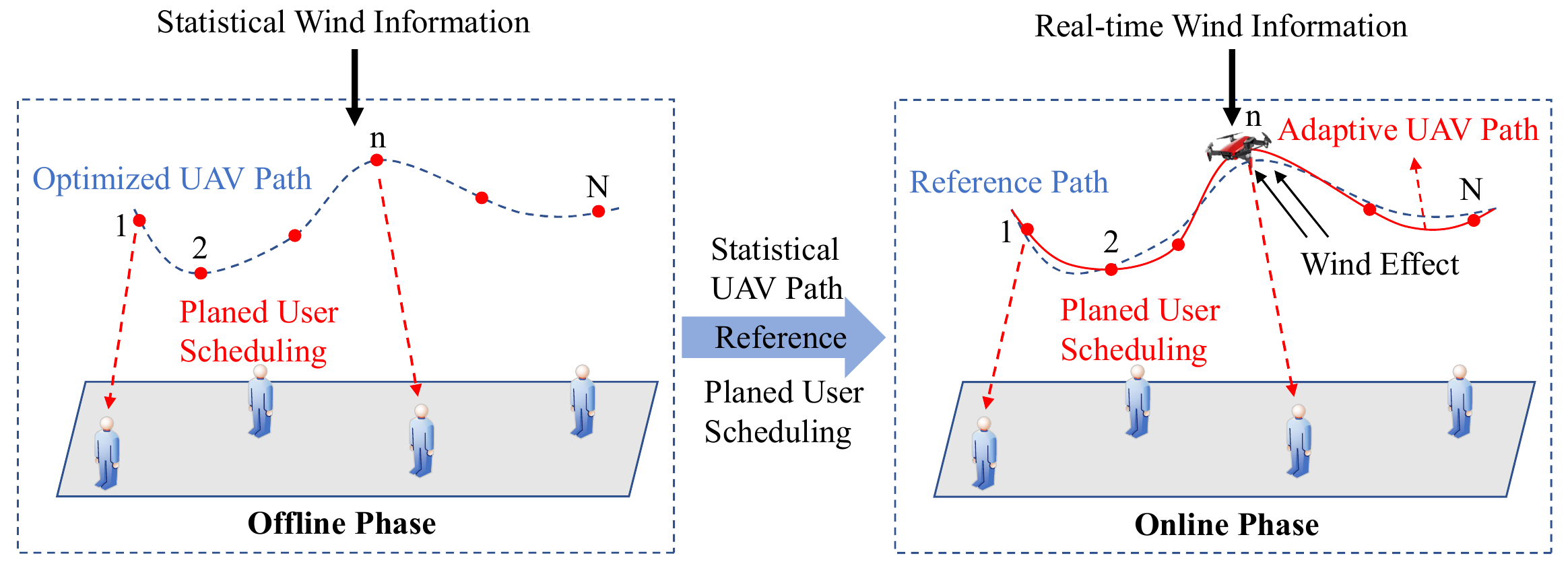}
    \caption{The proposed OBOA design for the UAV-enabled communication in wind}
    \label{OBOA}
\end{figure} 
    \item \textit{Online phase}: During the flight, to make the UAV adapt to the changeable real-time wind better and achieve a higher EE, we assume that the instantaneous wind information can be perfectly measured and collected by the UAV. Then, an OBOA design is proposed to minimize propulsion energy consumption slot by slot, by optimizing UAV instantaneous velocity with the real-time wind information and taking the obtained offline trajectory and user scheduling as references. Based on the OBOA design, the UAV can dynamically adjusts its velocity to further reduce the impact of the real-time wind and save more propulsion energy. In this case, the proposed OBOA design leads to a more energy-efficient solution to problem \eqref{problem_original} compared with the offline design.
\end{itemize}
\par In the following sections, we describe the two phases of the proposed OBOA design, including the offline design based on the given wind distribution and the OBOA design with real-time wind information and offline references.
\section{Offline Design}
In this section, to obtain a statistically suboptimal solution, we approximate the original problem \eqref{problem_original} into a tractable one, which is formulated as a stochastic program. Further, we decompose this problem into three subproblems and alternately optimize them to obtain an offline solution.
\subsection{Problem Approximation}
\par To maximize the EE of the UAV, we need to relax the objective function in \eqref{problem_original} as the ratio of the minimum average rate $R_{\rm min}$ from each GU to the upper bound of $P[n]$, denoted as $P^{\rm ub}[n]$. To this end, we tackle the non-convexity of $P_b[n]$, $P_i[n]$, and $m\|\mathbf{g}\|\|\mathbf{v}[n]\|\sin{\tau_c[n]}$ in $P[n]$. Specifically, to deal with $P_b[n]$, we introduce the slack variables $\mathbf{M}_1 = \{M_1[n]\}_{n=1}^N$ and $\mathbf{S}_1 = \{S_1[n]\}_{n=1}^N$ such that
$M_1[n] = \frac{1}{2\rho A}(\|m\mathbf{a}[n]-m\mathbf{g}\|+\frac{1}{2}\rho S_{\rm FP}\|(\mathbf{v}[n]-\mathbf{v}_w[n])\|^2) \ge \frac{\|\mathbf{T}[n]\|}{2\rho A} $ and $S_1[n] \ge (\frac{2}{c_T}\eta[n]+3\|\mathbf{v}[n]\|^2)\sqrt{\frac{2 \eta[n]}{c_T}}$, yielding
\begin{align}
\label{constraint_S1M1}
 \frac{c_T}{2}\frac{S_1^2[n]}{M_1[n]} \ge \left(\frac{2}{c_T}M_1[n]+3\|\mathbf{v}[n]\|^2\right)^2,\quad \forall n.
\end{align}
\par Similarly, for $P_i[n]$, we introduce the slack variables $\mathbf{M}_2 = \{M_2[n]\}_{n=1}^N$ and $\mathbf{S}_2 = \{S_2[n]\}_{n=1}^N$ satisfying $
M_2[n] \ge \sqrt{M_1^2[n]+\frac{\|\mathbf{v}[n]\|^4}{4}}-\frac{\|\mathbf{v}[n]\|^2}{2}$ and $S_2[n] \ge M_1[n]\sqrt{M_2[n]}$. Hence we have:
\begin{align}
\label{constraint_S2M2}
 \frac{S_2^2[n]}{M_2[n]} \ge M_1^2[n],\quad \forall n,
\end{align}
\begin{align}
\label{constraint_M1M2}
M_2[n]+\|\mathbf{v}[n]\|^2 \ge \frac{M_1^2[n]}{M_2[n]},\quad \forall n.
\end{align}
Note that the left-hand side (LHS) of \eqref{constraint_S1M1}, \eqref{constraint_S2M2} and \eqref{constraint_M1M2} are all convex and differentiable functions w.r.t. the slack variables, which must hold with equalities to obtain the optimal solution to problem \eqref{problem_original}. Otherwise $M_1[n]$ and $M_2[n]$ can be increased and so can $S_1[n]$ and $S_2[n]$ to further reduce the objective value of problem \eqref{problem_original}.
To convert the inequalities \eqref{constraint_S1M1}-\eqref{constraint_M1M2} into convex constraints, we apply the first-order Taylor expansion to obtain their linear lower bounds, at given local points denoted by $\mathbf{M}_1^t = \{M_1^t[n]\}_{n=1}^N$, $\mathbf{M}_2^t = \{M_2^t[n]\}_{n=1}^N$, $\mathbf{S}_1^t = \{S_1^t[n]\}_{n=1}^N$, and $\mathbf{S}_2^t = \{S_2^t[n]\}_{n=1}^N$ in the \emph{t}-th iteration, respectively. Then, we define $\mathbf{v}^t[n]=\frac{\mathbf{Q}^t[n+1]-\mathbf{Q}^t[n]}{\Delta}$, and have the following convex constraints:
\begin{align}
\label{constraint_S1M1_cvx}
\left(\frac{2}{c_T}M_1[n]+3\|\mathbf{v}[n]\|^2\right)^2 \le &\frac{c_T}{2}\left(\frac{(S_1^t[n])^2}{M_1^t[n]}+\frac{2S_1^t[n](S_1[n]-S^t_1[n])}{M_1^t[n]}\nonumber\right.\\
&\left.-\left(\frac{S_1^t[n]}{M_1^t[n]}\right)^2(M_1[n]-M_1^t[n])\right),\quad \forall n,
\end{align}
\begin{align}
\label{constraint_M1M2_cvx}
\frac{M_1^2[n]}{M_2[n]} \le M_2[n]-\|\mathbf{v}^t[n]\|^2+2(\mathbf{v}^t[n])^\mathrm{T}\mathbf{v}[n],\quad \forall n,
\end{align}
\begin{align}
\label{constraint_S2M2_cvx}
M_1^2[n] \le \frac{S_2^t[n]}{M_2^t[n]}+\frac{2S_2^t[n](S_2[n]-S_2^t[n])}{M_2^t[n]}-\left(\frac{S_2^t[n]}{M_2^t[n]}\right)^2(M_2[n]-M_2^t[n]),\quad \forall n.
\end{align}
Furthermore, for $P_c[n]=m\|\mathbf{g}\|\|\mathbf{v}[n]\|\sin{\tau_c[n]}$, we have
\begin{align}
m\|\mathbf{g}\|\|\mathbf{v}[n]\|\sin{\tau_c[n]} \le m\|\mathbf{g}\|\|\mathbf{v}[n]\|. 
\end{align}
Therefore, $P^{\rm ub}[n]$ can be rewritten as:
\begin{align}
P^{\rm ub}[n] = \frac{\delta\rho sA}{8}S_1[n]+2(1+c_f)\rho AS_2[n]+m\|\mathbf{g}\|\|\mathbf{v}[n]\|+\frac{1}{2}\rho S_{\rm FP}\|\mathbf{v}[n]-\mathbf{v}_w[n]\|^3.
\end{align}
\par We now tackle constraint \eqref{constraint_fairness_origin}. From \cite{you2020hybrid}, with given GUs' locations, the achievable rate in LoS state is typically much higher than that in NLoS state. To express the minimum expected rate, we take an approximation as:
\begin{align}
\label{rate expectation}
\mathbb{E}(R_k[n]) &=  P^L_k[n]R^L_k[n]+(1-P^L_k[n])R^N_k[n] \ge P^L_k[n]R^L_k[n]\nonumber\\
& = \left( A_3+\frac{A_4}{1+e^{(A_1+A_2\theta_k[n])}}\right)\log_2(1+\gamma_0 d_k^{-\alpha_L}[n]) \triangleq \overline{R}_k[n], \quad \forall n,k.
\end{align}
Then, to address the non-affine relationship between $\mathbf{Q}$ and $\overline{R}_k[n]$, combining \eqref{constraint_S1M1_cvx}-\eqref{rate expectation}, we reformulate problem \eqref{problem_original} into the following approximate form:
 \begin{subequations}\label{problem_relax}
 \begin{align} \underset{\mathbf{Q},\mathbf{A},\mathbf{\Theta},\mathbf{M}_1,\mathbf{M}_2,\mathbf{S}_1,\mathbf{S}_2,R_{\rm min}}{\max}\quad &\frac{R_{\rm min}}{\sum_{n=1}^{N}P^{\rm ub}[n]}\label{function_relax}\\
s.t.\quad 
& M_1[n] = \frac{1}{2\rho A}(\|m\mathbf{a}[n]-m\mathbf{g}\|+\frac{1}{2}\rho S_{\rm FP}\|(\mathbf{v}[n]-\mathbf{v}_w[n])\|^2),\quad \forall n,\label{constraint_M1}\\
& \theta_k[n] \le \frac{180}{\pi}\arctan\left(\frac{z[n]}{\|\mathbf{q}[n]-\mathbf{g}_k\|}\right), \quad \forall k,n,\label{constraint_theta}\\
& R_{\rm min}\le \sum_{n=1}^{N}a_k[n]\overline{R}_k[n],\quad \forall k,\label{constraint_fairness_lb}\\
&\eqref{constaint_v},\eqref{constaint_H},\eqref{constraint_a_0,1},\eqref{constraint_a<=1},\eqref{contraint_start_final},\eqref{constraint_S1M1_cvx},\eqref{constraint_M1M2_cvx},\eqref{constraint_S2M2_cvx},\nonumber
\end{align}
\end{subequations}
where $\mathbf{\Theta}=\{\theta_k[n]\}_{n=1}^N, \forall k$. Note that \eqref{constraint_fairness_lb} is still a non-convex constraint due to the coupled horizontal and vertical trajectory variables of $\theta_k[n]$ in $\overline{R}_k[n]$.
\subsection{Proposed Offline Design on SP}
To average the effect of $\mathbf{v}_w$, a stochastic fractional program yielding a statistical maximum EE of the UAV is formulated as:
\begin{subequations}\label{problem_relax_SP}
 \begin{align}
\underset{\mathbf{Q},\mathbf{A},\mathbf{\Theta},\mathbf{M}_1,\mathbf{M}_2,\mathbf{S}_1,\mathbf{S}_2,R_{\rm min}}{\max}\quad &\frac{R_{\rm min}}{\sum_{n=1}^{N}(P^{\rm ub}_b[n]+P^{\rm ub}_i[n]+P^{\rm ub}_c[n]+\frac{1}{2}\rho S_{\rm FP}\mathbb{E}(\|\mathbf{v}[n]-\mathbf{v}_w[n]\|^3))}\label{function_relax_SP}\\
s.t.\quad 
&M_1[n] = \frac{1}{2\rho A}(\|m\mathbf{a}[n]-m\mathbf{g}\|+\frac{1}{2}\rho S_{\rm FP}\mathbb{E}\|(\mathbf{v}[n]-\mathbf{v}_w[n])\|^2),\quad \forall n,\label{constraint_M1_SP}\\
&\eqref{constaint_v},\eqref{constaint_H},\eqref{constraint_a_0,1},\eqref{constraint_a<=1},\eqref{contraint_start_final},\eqref{constraint_S1M1_cvx},\eqref{constraint_M1M2_cvx},\eqref{constraint_S2M2_cvx},\eqref{constraint_theta},\eqref{constraint_fairness_lb},\nonumber
\end{align}
\end{subequations}
where $P^{\rm ub}_b[n] = \frac{\delta\rho sA}{8}S_1[n]$, $P^{\rm ub}_i[n] = 2(1+c_f)\rho AS_2[n]$ and $P^{\rm ub}_c[n] = m\|\mathbf{g}\|\|\mathbf{v}[n]\|$. Note that $\mathbf{v}_w[n]$ is a function of $v_{\rm ref}[n]$, $\beta[n]$ and $z[n]$ as shown in \eqref{v_w}, $\mathbf{v}[n] - \mathbf{v}_w[n]$ is convex w.r.t. $\mathbf{z}[n]$ and also $\mathbf{Q}[n]$ when $p\le1$. Since the expectation preserves convexity, $\mathbb{E}(\|\mathbf{v}[n]-\mathbf{v}_w[n]\|^3)$ in $P^{\rm ub}[n]$ and constraint \eqref{constraint_M1_SP} are still convex.
\par We use the Monte Carlo sample average approximation (MCSAA), i.e., a method of SP \cite{shapiro2008stochastic} to deal with the expectations in problem \eqref{problem_relax_SP}. Specifically, we take an approximation as $\mathbb{E}\{f(\mathbf{v}[n],\mathbf{v}_w[n])\} \approx \frac{1}{S}\sum_{i=1}^{S}f(\mathbf{v}[n],\mathbf{v}^i_w[n])$, where $f$ is a general function of the optimization variable $\mathbf{v}[n]$ and the random variable $\mathbf{v}_w[n]$; $\mathbf{v}^i_w[n]$ is the $i$-th sample generated from the given random distribution of the wind; $S$ denotes a finite set of random samples. Clearly, the above approximation becomes accurate for a sufficiently large $S$.
\par To tackle the non-convexity of binary $\mathbf{A}$ and \eqref{constraint_fairness_lb}, we next decompose problem \eqref{problem_relax_SP} into three subproblems, where the user scheduling $\mathbf{A}$, the horizontal trajectory $\mathbf{q}$ and the vertical trajectory $\mathbf{z}$ are optimized in an alternate fashion. 
\par \subsubsection{User Scheduling Optimization}For given $\hat{\mathbf{Q}}=\{(\hat{\mathbf{q}}[n]^{\mathrm{T}},\hat{z}[n])\}_{n=1}^N$, the subproblem is equivalent to maximize $R_{\rm min}$ in \eqref{function_relax_SP}. To make the subproblem tractable, we relax the integer variables in \eqref{constraint_a_0,1} into continuous variables by considering that all $a_k[n]$ are continuous between 0 and 1, yielding the following subproblem:
 \begin{subequations}\label{subproblem_alpha_relax}
 \begin{align}
 \underset{\mathbf{A},R_{\rm min}}{\max}\quad &R_{\rm min}\label{function_alpha_relax}\\
s.t.\quad&0\le a_k[n]\le 1,\quad \forall n,k,\label{constraint_a_relax_0,1}\\
&\eqref{constraint_a<=1}, \eqref{constraint_fairness_lb}.\nonumber
\end{align}
\end{subequations}
Since subproblem \eqref{subproblem_alpha_relax} is a linear program, it can be solved efficiently by CVX.
\par \subsubsection{UAV Horizontal Trajectory Optimization}Given $\hat{\mathbf{A}}$ and $\hat{\mathbf{z}}$, the UAV horizontal trajectory $\mathbf{q}$ can be optimized by solving the following subproblem:
\begin{subequations}\label{subproblem_q}
 \begin{align}
 \underset{\mathbf{q},\mathbf{\Theta},\mathbf{M}_1,\mathbf{M}_2,\mathbf{S}_1,\mathbf{S}_2,R_{\rm min}}{\max}\quad &\frac{R_{\rm min}}{\sum_{n=1}^{N}(P^{\rm ub}_b[n]+P^{\rm ub}_i[n]+P^{\rm ub}_c[n]+\frac{1}{2}\rho S_{\rm FP}\mathbb{E}(\|\mathbf{v}[n]-\mathbf{v}_w[n]\|^3))}\label{function_q}\\
s.t.\quad &\mathbf{Q[n]} = (\mathbf{q}^\mathrm{T}[n],\hat{z}[n]), \quad \forall n,\label{constraint_Q_q}\\
&\theta_k[n] \le \frac{180}{\pi}\arctan\left(\frac{\hat{z}[n]}{\|\mathbf{q}[n]-\mathbf{g}_k\|}\right), \quad \forall k,n,\label{constraint_theta_q}\\
& R_{\rm min}\le \sum_{n=1}^{N}\hat{a}_k[n]\overline{R}_k[n],\quad \forall k,\label{constraint_fairness_lb_q}\\
&\eqref{constaint_v},\eqref{contraint_start_final},\eqref{constraint_S1M1_cvx},\eqref{constraint_M1M2_cvx},\eqref{constraint_S2M2_cvx},\eqref{constraint_M1_SP}.\nonumber
\end{align}
\end{subequations}
To solve this subproblem, from [\citenum{you2020hybrid}, \textit{Lemma 1}], we see that $\overline{R}_k[n]$ is a convex function w.r.t. $(1+e^{-(A_1+A_2\theta_k[n])})$ and $(z[n]^2+\|\mathbf{q}[n]-\mathbf{g}_k\|^2)$. Thus, we obtain a lower bound of $\overline{R}_k[n]$, by using the first-order Taylor expansion for any local UAV horizontal trajectory $\mathbf{Q}^t[n] = ((\mathbf{q}^t[n])^\mathrm{T},\hat{z}[n])$ in the $t$-th iteration:
\begin{align}
\overline{R}_k[n] \ge &(A_3+\frac{A_4}{\xi^t_k[n]})\log_2\left(1+\frac{\gamma_0}{(\phi^t_k[n])^{\frac{1}{2}\alpha_L}}\right)-\frac{A_4\log_2e}{(\xi^t_k[n])^2}\ln\left(1+\frac{\gamma_0}{(\phi^t_k[n])^{\frac{1}{2}\alpha_L}}\right)\nonumber\\
 & \times (e^{-u_k[n]}-e^{-u^t_k[n]})- (A_3+\frac{A_4}{\xi^t_k[n]})\left(\frac{\gamma_0\alpha_L}{2((\phi^t_k[n])^{\frac{1}{2}\alpha_L}+\gamma_0)\phi^t_k[n]}\right) \nonumber\\
& \times (\|\mathbf{q}[n]-\mathbf{g}_k\|^2-\|\mathbf{q}^t[n]-\mathbf{g}_k\|^2) \triangleq  \overline{R}^{\rm lb,\mathbf{q}}_k[n],
\end{align}where $u_k[n]\triangleq A_1 + A_2\theta_k[n]$, $u^t_k[n]\triangleq A_1 + A_2\theta^t_k[n]$, $\xi^t_k[n] = 1+e^{-(A_1 + A_2\theta^t_k[n])}$ and $\phi^t_k[n] = \hat{z}[n]^2+\|\mathbf{q}^t[n]-\mathbf{g}_k\|^2$. Similarly, for non-concave term $\arctan\left(\frac{\hat{z}[n]}{\|\mathbf{q}[n]-\mathbf{g}_k\|}\right)$ w.r.t. $\mathbf{q}[n]$, since $\arctan(1/x)$ is convex for $x>0$, we obtain a lower bound as
\begin{align}
\frac{180}{\pi}\arctan\left(\frac{\hat{z}[n]}{\|\mathbf{q}[n]-\mathbf{g}_k\|}\right) &\ge \frac{180}{\pi}\theta_k^{\rm lb}[n] \ge \theta_k[n],
\end{align}
where 
\begin{align}
\theta_k^{\rm lb}[n] \triangleq \arctan\left(\frac{\hat{z}[n]}{\|\mathbf{q}^t[n]-\mathbf{g}_k\|}\right)-\frac{\hat{z}[n]}{\phi^t_k[n]}(\|\mathbf{q}[n]-\mathbf{g}_k\|-\|\mathbf{q}^t[n]-\mathbf{g}_k\|). 
\end{align}
Therefore, subproblem \eqref{subproblem_q} can be recast as
\begin{subequations}\label{subproblem_q_relax}
 \begin{align}
 \underset{\mathbf{q},\mathbf{\Theta},\mathbf{U},\mathbf{M}_1,\mathbf{M}_2,\mathbf{S}_1,\mathbf{S}_2,R_{\rm min}}{\max}\quad &\frac{R_{\rm min}}{\sum_{n=1}^{N}(P^{\rm ub}_b[n]+P^{\rm ub}_i[n]+P^{\rm ub}_c[n]+\frac{1}{2}\rho S_{\rm FP}\mathbb{E}(\|\mathbf{v}[n]-\mathbf{v}_w[n]\|^3))}\label{function_q_relax}\\
s.t.\quad & R_{\rm min}\le \sum_{n=1}^{N}\hat{a}_k[n]\overline{R}^{\rm lb,\mathbf{q}}_k[n],\quad \forall k,\label{constraint_fairness_lb_q_relax}\\
& \theta_k[n] \le \frac{180}{\pi}\theta^{\rm lb}_k[n],\quad \forall k,n,\label{constraint_theta_q_relax}\\
& u_k[n] = A_1 + A_2\theta_k[n],\quad \forall k,n,\label{constraint_u}\\
&\eqref{constaint_v},\eqref{contraint_start_final},\eqref{constraint_S1M1_cvx},\eqref{constraint_M1M2_cvx},\eqref{constraint_S2M2_cvx},\eqref{constraint_M1_SP},\eqref{constraint_Q_q},\nonumber
\end{align}
\end{subequations}
where $\mathbf{U}=\{u_k[n]\}_{n=1}^N, \forall k$. Then, subproblem \eqref{subproblem_q_relax} is a quasi-convex optimization problem, which can be solved efficiently via successive convex approximation (SCA) and Dinkelbach’s method \cite{shen2018fractional}.
\par \subsubsection{UAV Vertical Trajectory Optimization}Given $\hat{\mathbf{A}}$ and $\hat{\mathbf{q}}$, the UAV vertical trajectory $\mathbf{z}$ can be optimized by solving the following subproblem:
\begin{subequations}\label{subproblem_z}
 \begin{align}
\underset{\mathbf{z},\mathbf{\Theta},\mathbf{M}_1,\mathbf{M}_2,\mathbf{S}_1,\mathbf{S}_2,R_{\rm min}}{\max}\quad &\frac{R_{\rm min}}{\sum_{n=1}^{N}(P^{\rm ub}_b[n]+P^{\rm ub}_i[n]+P^{\rm ub}_c[n]+\frac{1}{2}\rho S_{\rm FP}\mathbb{E}(\|\mathbf{v}[n]-\mathbf{v}_w[n]\|^3))}\label{function_z}\\
s.t.\quad &\mathbf{Q[n]} = (\hat{\mathbf{q}}^\mathrm{T}[n],z[n]), \quad \forall n,\label{constraint_Q_z}\\
&\theta_k[n] \le \frac{180}{\pi}\arctan\left(\frac{z[n]}{\|\hat{\mathbf{q}}[n]-\mathbf{g}_k\|}\right), \quad \forall k,n,\label{constraint_theta_z}\\
& R_{\rm min}\le \sum_{n=1}^{N}\hat{a}_k[n]\overline{R}_k[n],\quad \forall k,\label{constraint_fairness_lb_z}\\
&\eqref{constaint_v},\eqref{constaint_H},\eqref{contraint_start_final},\eqref{constraint_S1M1_cvx},\eqref{constraint_M1M2_cvx},\eqref{constraint_S2M2_cvx},\eqref{constraint_M1_SP}.\nonumber
\end{align}
\end{subequations}
Note that subproblem \eqref{subproblem_z} is similar to \eqref{subproblem_q}, yielding the following subproblem:
\begin{subequations}\label{subproblem_z_relax}
 \begin{align}
 \underset{\mathbf{z},\mathbf{U},\mathbf{\Theta},\mathbf{M}_1,\mathbf{M}_2,\mathbf{S}_1,\mathbf{S}_2,R_{\rm min}}{\max}\quad &\frac{R_{\rm min}}{\sum_{n=1}^{N}(P^{\rm ub}_b[n]+P^{\rm ub}_i[n]+P^{\rm ub}_c[n]+\frac{1}{2}\rho S_{\rm FP}\mathbb{E}(\|\mathbf{v}[n]-\mathbf{v}_w[n]\|^3))}\label{function_z_relax}\\
s.t.\quad & R_{\rm min}\le \sum_{n=1}^{N}\hat{a}_k[n]\overline{R}^{\rm lb,\mathbf{z}}_k[n],\quad \forall k,\label{constraint_fairness_lb_z_relax}\\
&\eqref{constaint_v},\eqref{constaint_H},\eqref{contraint_start_final},\eqref{constraint_S1M1_cvx},\eqref{constraint_M1M2_cvx},\eqref{constraint_S2M2_cvx},\eqref{constraint_M1_SP},\eqref{constraint_u},\eqref{constraint_Q_z},\eqref{constraint_theta_z},\nonumber
\end{align}
\end{subequations}
where $\overline{R}^{\rm lb,\mathbf{z}}_k[n]$ is the lower bound of $\overline{R}_k[n]$ for any local UAV vertical trajectory $\mathbf{Q}^t[n] = (\hat{\mathbf{q}}[n])^\mathrm{T},z^t[n])$ in $t$-th iteration, which is similar to $\overline{R}^{\rm lb,\mathbf{q}}_k[n]$. Now all constraints of subproblem \eqref{function_z_relax} are convex, and it can be solved similarly as in \eqref{subproblem_q_relax}.
\begin{algorithm}[t]
\caption{ Proposed Offline Design on SP} 
\label{Alogrithm_online} 
\begin{algorithmic}[1] 
\REQUIRE $\mathbf{g}_k, k \in \mathcal{K}$, wind samples $\mathbf{v}^i_w[n], n\in \mathcal{N}, i\in \mathcal{S}$, convergence threshold $\epsilon_1$ and $\epsilon_2$.
\STATE {\textbf{Initialization:}} Let $\mathbf{q}^0,\mathbf{z}^0$, $\mathbf{A}^0$ be feasible for problem \eqref{problem_relax_SP}; Set iteration index $r=0$;
\REPEAT
\STATE Solve problem \eqref{subproblem_alpha_relax} with given $\mathbf{q}^r$ and $\mathbf{z}^r$, and denote $\mathbf{A}^{r+1}$ by the optimal solution;
\STATE Solve problem \eqref{subproblem_q_relax} based on SCA and Dinkelbach's method with given $\mathbf{A}^{r+1}$ and $\mathbf{z}^r$, until its computed objective value converges within $\epsilon_1 > 0$, and denote $\mathbf{q}^{r+1}$ by the optimal solution;
\STATE Solve problem \eqref{subproblem_z_relax} based on the same methods in step 4 with given $\mathbf{A}^{r+1}$ and $\mathbf{q}^{r+1}$, until its computed objective value converges within $\epsilon_1 > 0$, and denote $\mathbf{z}^{r+1}$ by the optimal solution;
\STATE Update $r = r+1$.
\UNTIL The computed objective value of problem \eqref{problem_relax_SP} converges
within $\epsilon_2 > 0$.
\ENSURE $\mathbf{Q[n]} = (\mathbf{q}^\mathrm{T}[n],z[n]), \forall n, \mathbf{A}$.%算法的输出：Output
\end{algorithmic}
\end{algorithm}
\subsection{Overall Algorithm and Computational Complexity}
From the analysis in preceding subsections, the proposed offline design on SP (summarized in Algorithm 1) provides the suboptimal offline trajectory and user scheduling policy of problem \eqref{problem_relax_SP} by alternately optimizing $\mathbf{A}$, $\mathbf{q}$ and $\mathbf{z}$. Note that the user scheduling obtained by solving subproblem \eqref{subproblem_alpha_relax} is continuous. To compute the objective value of problem \eqref{problem_relax_SP} accurately, we rounding the optimized user scheduling into a binary solution after Algorithm 1 is completed \cite{wu2018joint}. {Note that the decomposed three subproblems satisfy that their optimal values all serve as a lower bound of the optimal value of problem \eqref{problem_relax_SP}, and the values are non-decreasing after each iteration. In addition, it is clear that the maximum objective value of problem \eqref{problem_relax_SP} is upper bounded by a finite value, which guarantees the convergence of Algorithm 1 as analyzed in \cite{ben2001lectures}.} For the horizontal and vertical trajectory optimizations, both subproblems involve $2N$ $l_2$-norm terms with samples size $S$. Then, since all subproblems above are solved based on the standard interior-point method, and based on the complexity analysis method in \cite{wang2014outage}, the required computational complexity of Algorithm 1 is thus $\mathcal{O}\left(((KN+8N)^{3.5}+K^{1.5}S^2N^{2.5})\log(1/\epsilon_1)\log(1/\epsilon_2)\right)$.
\section{Offline-Based Online Adaptive Design}
The proposed offline design achieves an expected EE by jointly optimizing 3D UAV trajectory and user scheduling over the whole flight period, based on the given wind statistics. Although we obtain a statistically favorable suboptimal solution to problem \eqref{problem_original}, the practical wind affecting UAV flight may be significantly different from the predicted statistics, e.g., the speed and direction of the real-time wind have large standard deviations, and change dramatically. In this case, the obtained offline solution is not energy-efficient enough. 
\par In this section, the OBOA design with low complexity is proposed to make the UAV better adapt to changeable real-time wind. Although the offline solution is only statistically favorable, it still provides a suboptimal trajectory design and user scheduling policy considering the real-time wind. Thus, in the online phase, to take full advantage of the offline design and further improve the EE, we denote the obtained trajectory and user scheduling policy in the offline phase as $\mathbf{Q}^{\rm off}$ and $\mathbf{A}^{\rm off}$, respectively, which are used as references for the OBOA design. In practice, the complexity of the online adaptive design must be relatively low that we can execute the algorithm within the given time slot. From the analysis of the offline design, \eqref{constraint_fairness_lb_q_relax} and other communication related constraints lead to high computational complexity so that the proposed design is difficult to implement in practice. Since the expected achievable rate is only related to the distance and scheduling between the UAV and each GU, we make the following assumptions: i) Under reasonable trajectory deviation, expected achievable rate of the UAV with the online design is slightly lower than the rate with the offline design. ii) The UAV can measure and collect the wind data in real time, including the speed and direction, and we ignore the collection delay of sensors deployed on the UAV.
\par Based on the assumptions above, the OBOA design can be implemented by only optimizing the real-time velocity of the UAV to minimize its propulsion energy consumption in each slot, regardless of the constraints related to the communication. As a result, an online adaptation problem at each slot \textit{n} can be modified as 
\begin{subequations}\label{problem_online}
 \begin{align}
\underset{\mathbf{v[n]}}{\min}\quad & {P[n]}\label{function_online}\\
s.t.\quad& \mathbf{Q}[n+1] = \mathbf{Q}[n] + \mathbf{v}[n]\Delta,\label{constraint_online_Q}\\
& \|\mathbf{Q}[n+1]-\mathbf{Q}^{\rm off}[n+1]\| \le \varepsilon _Q,\label{constraint_online_distance}\\
& \|\mathbf{v}[n]-\mathbf{v}^{\rm off}[n]\| \le \varepsilon_v,\label{constraint_online_v}\\
& \|\mathbf{Q}[n+1]-\mathbf{Q}_{\rm F}\| \le \|\mathbf{Q}^{\rm off}[n+1]-\mathbf{Q}_{\rm F}\|,\label{constraint_online_final}
\end{align}
\end{subequations}
where $\hat{\mathbf{v}}_w[n]$ denotes the wind measured by the UAV in real-time in time slot $n$. $\mathbf{Q}[n]$ represents the current position of the UAV, i.e., 3D position obtained by the online optimization in the last time slot. $\varepsilon_Q$ and $\varepsilon_v$ are two tolerance parameters to restrict the extents of deviation from the offline trajectory $\mathbf{Q}^{\rm off}$ and velocity $\mathbf{v}^{\rm off}$, respectively, and the larger tolerances mean that the UAV has more space to adjust. {\eqref{constraint_online_final} ensures that in the whole online adaptation process, the UAV can dynamically adjust flight direction and control its speed $\|\mathbf{v}[n]\|$ properly to avoid flying at the maximum-endurance speed defined in \cite{zeng2019energy} in each slot, so that $\|\mathbf{v}[n]\|$ does not deviate much from  $\|\mathbf{v}^{\rm off}[n]\|$.} Problem \eqref{problem_online} is difficult to solve due to the non-convexity of $P[n]$. To tackle this, we minimize its upper bound by the approximation analyzed in Section \Rmnum{5}-A. Then, problem \eqref{problem_online} is rewritten as:
\begin{subequations}\label{problem_online_sca}
 \begin{align}
\underset{\mathbf{v[n]},M_1[n],M_2[n],S_1[n],S_2[n]}{\min}\quad & {P^{\rm ub}_b[n]+P^{\rm ub}_i[n]+P^{\rm ub}_c[n]+\frac{1}{2}\rho S_{\rm FP}\|\mathbf{v}[n]-\hat{\mathbf{v}}_w[n]\|^3}\label{function_online_sca}\\
s.t.\quad& M_1[n] = \frac{1}{2\rho A}(\|m\mathbf{a}[n]-m\mathbf{g}\|+\frac{1}{2}\rho S_{\rm FP}\|(\mathbf{v}[n]-\hat{\mathbf{v}}_w[n])\|^2) ,\\
&\eqref{constraint_S1M1_cvx},\eqref{constraint_M1M2_cvx},\eqref{constraint_S2M2_cvx},\eqref{constraint_online_Q},\eqref{constraint_online_distance},\eqref{constraint_online_v},\eqref{constraint_online_final},\nonumber
\end{align}
\end{subequations}
which can be solved efficiently by CVX. {Note that the proposed adaptive design requires to perform just once the SCA iteration in each given time slot $n$ (summarized in Algorithm 2), and the optimization variable dimension of problem \eqref{problem_online_sca} is fixed at 7 (i.e., a three-dimensional vector $\mathbf{v}[n]$ and four one-dimensional slack variables). Since solving problem \eqref{problem_online_sca} in each slot by using the CVX solver is based on the standard interior-point method, its computational complexity scale as $\mathcal{O}\left((7)^{3.5}\log(1/\epsilon_1)\right)$, which is practically affordable for common UAV-mounted computer. Then, the totally required computational complexity of Algorithm 2 in the whole flight duration is $\mathcal{O}\left((N-1)(7)^{3.5}\log(1/\epsilon_1)\right)$, which do not increase with the number of GUs $K$. Therefore, the proposed OBOA design can be implemented practically within a given time slot}\footnote{The average running time of the OBOA design is about 0.5 s for each time slot. The result is obtained by using CVX and running with 3.6 GHz CPU and 8 GB RAM memory.}.
\begin{algorithm}[t]
\caption{ Proposed OBOA Design} 
\label{Alogrithm_offline} 
\begin{algorithmic}[1] 
\REQUIRE $\varepsilon_Q$, $\varepsilon_v$, offline references $\mathbf{Q}^{\rm off}$, $\mathbf{v}^{\rm off}$ and $\mathbf{A}^{\rm off}$, convergence threshold $\epsilon_3$.
\STATE {\textbf{Initialization:}} Set $\mathbf{v}^0[n] = \mathbf{v}^{\rm off}[n]$, $\mathbf{Q}[1] = \mathbf{Q}_{{\rm I}}$, $\mathbf{Q}[N] = \mathbf{Q}_{{\rm F}}$ and SCA iteration index $t=0$;
\FOR {$n = 1$ to $N-1$}
\STATE Collect the instantaneous wind information $\hat{\mathbf{v}}_w[n]$;
\STATE Solve problem \eqref{problem_online_sca} based on SCA with $\mathbf{Q}[n]$, $\mathbf{Q}^{\rm off}[n+1]$, $\mathbf{v}^{\rm off}[n]$ and $\hat{\mathbf{v}}_w[n]$, until its computed objective value converges within $\epsilon_3 > 0$, and denote $\mathbf{v}^{\rm opt}[n]$ by the optimal solution;
\STATE Obtain $\mathbf{Q}[n+1] = \mathbf{Q}[n] + \mathbf{v}^{\rm opt}[n]\Delta$.
\ENDFOR
\ENSURE $\mathbf{Q}, \mathbf{v}^{\rm opt}$.
\STATE UAV flies following the obtained $\mathbf{Q}$ and $\mathbf{v}^{\rm opt}$, and communicates to each GU with the scheduling policy $\mathbf{A}^{\rm off}$.
\end{algorithmic}
\end{algorithm}
\par{ It is worth mentioning that, as shown in step 7 of Algorithm 2, the UAV flies following the obtained $\mathbf{Q}$ and $\mathbf{v}^{\rm opt}$, and communicates to each GU with the scheduling policy $\mathbf{A}^{\rm off}$. Since we consider the real-time wind affecting UAV flight in the online adaptive optimization, we also consider the practical channel model depending on the real environment instead of the probability, to obtain the achievable rate and also the EE of the UAV more accurately in the online phase. Specifically, the achievable rate $R_k[n]$ is computed by $R^L_k[n]$ when the UAV-GU link is conditioned on the LoS state, and computed by $R^N_k[n]$ when buildings block the UAV-GU link. The practical channel in the LoS/NLoS states is based on a concrete Manhattan city model\cite{you20193d}, as shown in the following numerical results.}
 \begin{table}[htb] 
\small
\begin{center}   
\caption{Simulation Parameters}  
\label{table} 
\begin{tabular}{|m{4cm}<{\centering}|m{5.5cm}<{\centering}|m{5.5cm}<{\centering}|}   
\hline   \textbf{Parameter} & \textbf{Description} & \textbf{Value} \\ 
\hline  $T_0$, $\Delta$  &  Flight duration and time slot &  150 s, 1 s\\
\hline   $\mathbf{Q}_{\rm I},\mathbf{Q}_{\rm F}$ & Initial and final position & $(0,500,100)$ m, $(1000,500,100)$ m \\ 
\hline   $V^{\rm max}_H,V^{\rm max}_V$ & Maximum horizontal and vertical flight speeds & $40$ m/s, $20$ m/s  \\  
\hline   $H_{\rm max}$, $H_{\rm min}$ & Maximum and minimum flight altitudes & $300$ m, $50$ m  \\     
\hline   $K$, $\mathbf{g}_k$ & Number of GUs and corresponding horizontal coordinates  & 4, $(100,300)$ m, $(500,800)$ m, $(500,200)$ m, $(900,600)$ m \\ 
\hline   $A_1-A_4$ & Probabilistic LoS channel parameters & -1, 0.05, 0.1, 0.9  \\
\hline   $\rho_0$, $\mu_0$, $\Gamma$, $\alpha_L$, $\alpha_N$, $\sigma^2$, $B$, $P_0$ & Communication parameters & -60 dB, -20 dB, 8.2 dB, 2.5, 5, -110 dBm, 1 MHz, 0.1 W  \\
\hline   $m$, $\|\mathbf{g}\|$ & Weight of the UAV and gravitational acceleration & 2 $\rm kg$, 9.8 $\rm m/s^2$  \\
\hline   $\rho$, $A$, $\delta$, $s$, $S_{FP}$, $c_T$, $c_f$ & UAV aerodynamics parameters & 1.225 $\rm kg/m^3$, 0.79 $\rm m^2$, 0.012, 0.1, 0.01 $\rm m^2$, 0.3, 0.13  \\
\hline   $h_{\rm ref}$ & Reference altitude & 50 m \\
\hline   $p$ & Empirical exponent of the wind modeling & 0.5 \\
\hline   $\varepsilon_v$ & Speed tolerance & 20 m/s \\
\hline   {$\epsilon_1$, $\epsilon_2$, $\epsilon_3$, $S$} & {Convergence thresholds and size of random wind samples set} & {$10^{-3}$, 300} \\
\hline
\end{tabular}   
\end{center}   
\end{table}
 \section{Numerical Results}
 \par {In this section, numerical simulations are conducted to verify the effectiveness of the proposed offline and OBOA designs compared with the benchmark scheme (denoted as Windless). The trajectory and user scheduling of windless scheme are optimized by following the similar steps in Algorithm 1 considering a windless environment, i.e., solving the corresponding problems in Algorithm 1 based on the GPECM with $\|\mathbf{v}_w\| = 0$.} The parameter setting for the simulations are summarized in Table \ref{table}. Note that the parameters about communication and probabilistic LoS channel are similar to those in [\citenum{you2020hybrid}, \citenum{duo2020anti}], and the UAV aerodynamics parameters are set similarly as in [\citenum{zeng2019energy}, \citenum{9847346}, \citenum{cai2022resource}].
 
 \subsection{Performance Analysis for Offline Design}
\begin{figure}[htbp]
    \subfloat{
        \includegraphics[width=0.48\linewidth]{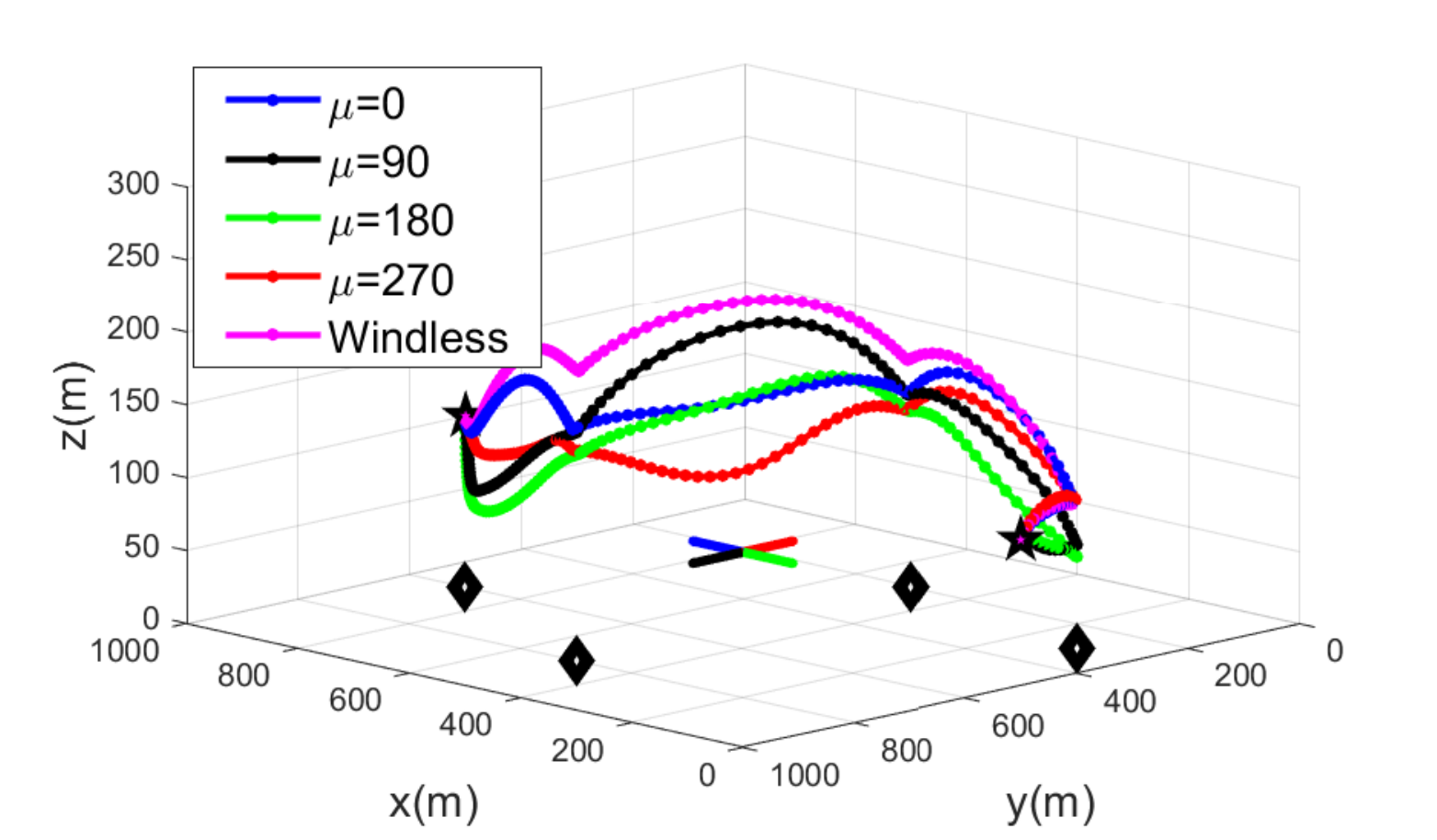}  
    }
    \subfloat{
        \includegraphics[width=0.48\linewidth]{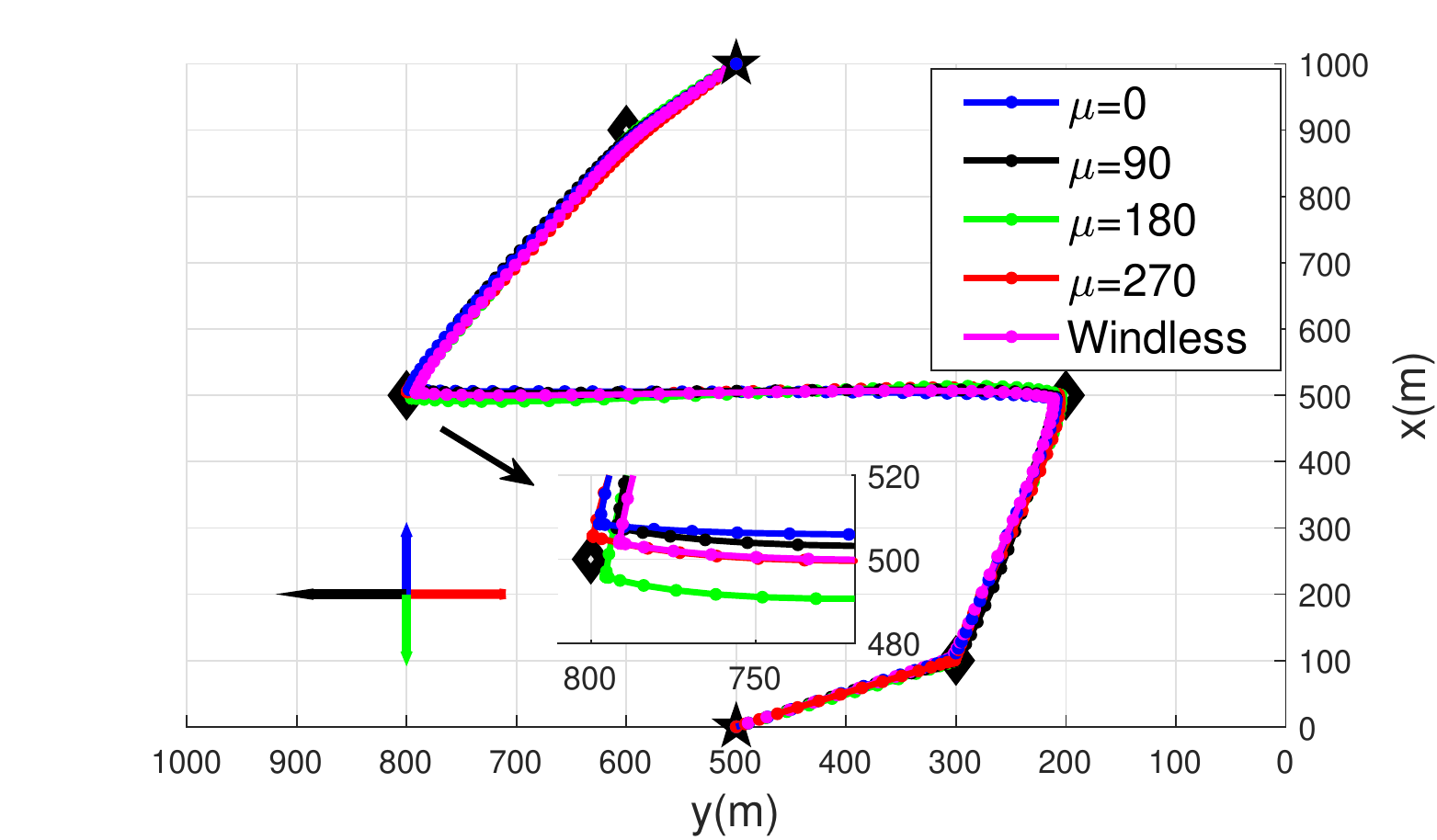}
    }
 \caption{ UAV trajectories based on different $\mu$ ($\kappa = 20$, $\lambda = 10$, $c = 10$)}
 \label{fig_traj_offline_theta}
\end{figure}

\begin{figure}[htbp]
    \subfloat{
        \includegraphics[width=0.48\linewidth]{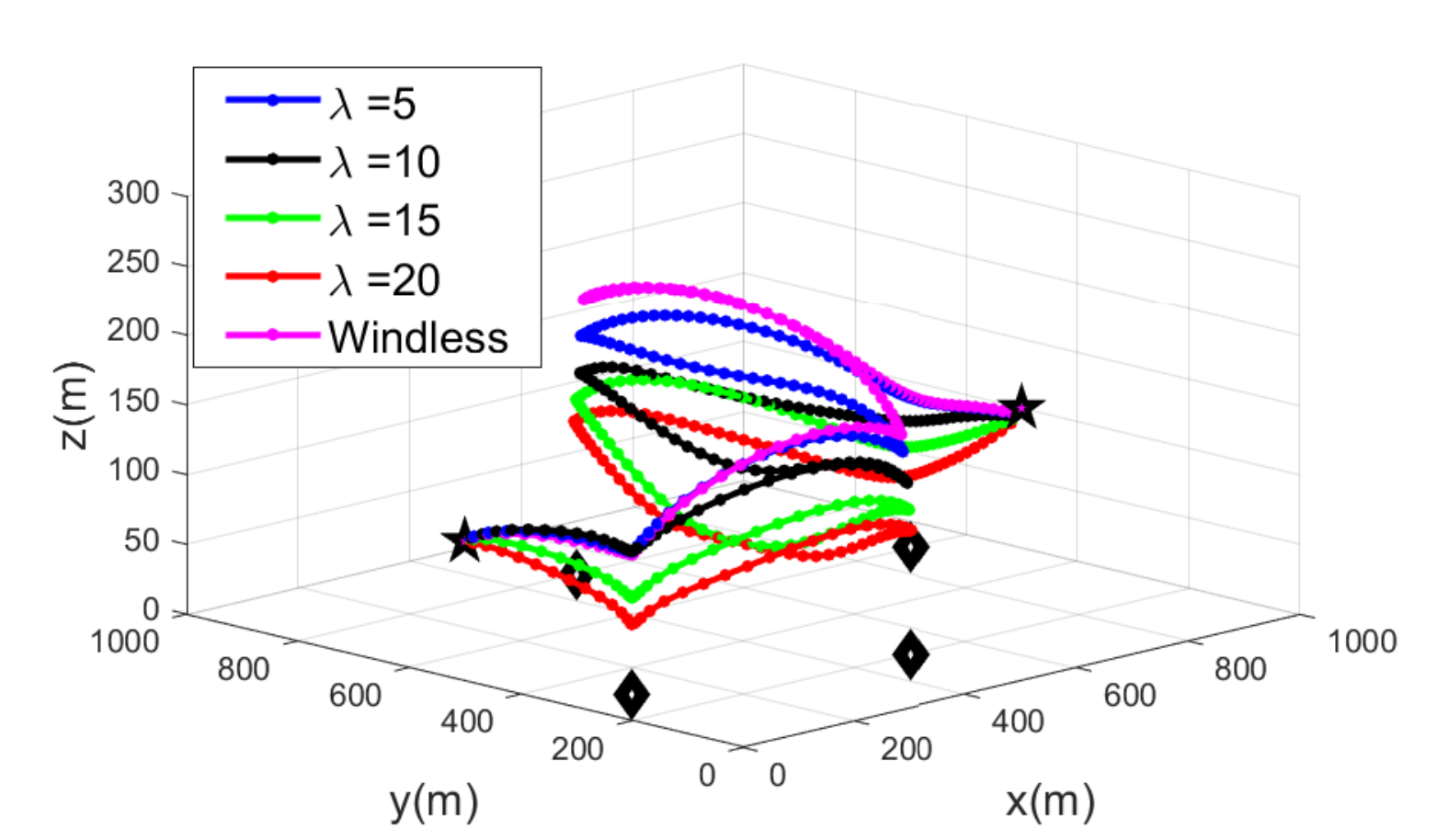}
    }
    \subfloat{
        \includegraphics[width=0.48\linewidth]{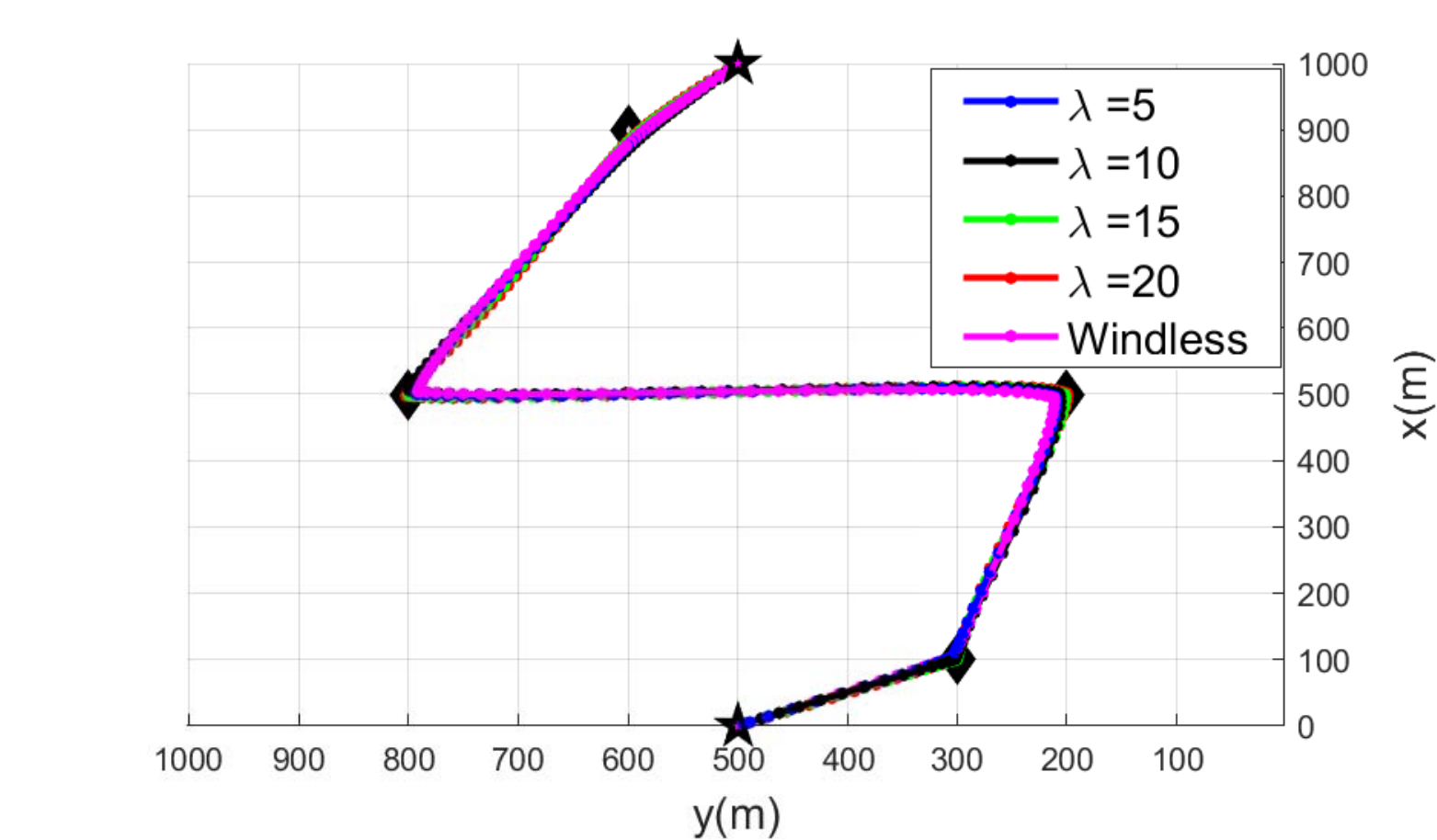}
    }
 \caption{UAV trajectories based on different $\lambda$ ($\mu = 270$, $\kappa = 20$, $c = 10$)}
 \label{fig_traj_offline_speed}
\end{figure}

\begin{figure}[htbp]
\centering
\begin{minipage}[t]{0.47\textwidth}
    \centering
    \includegraphics[width=1\textwidth]{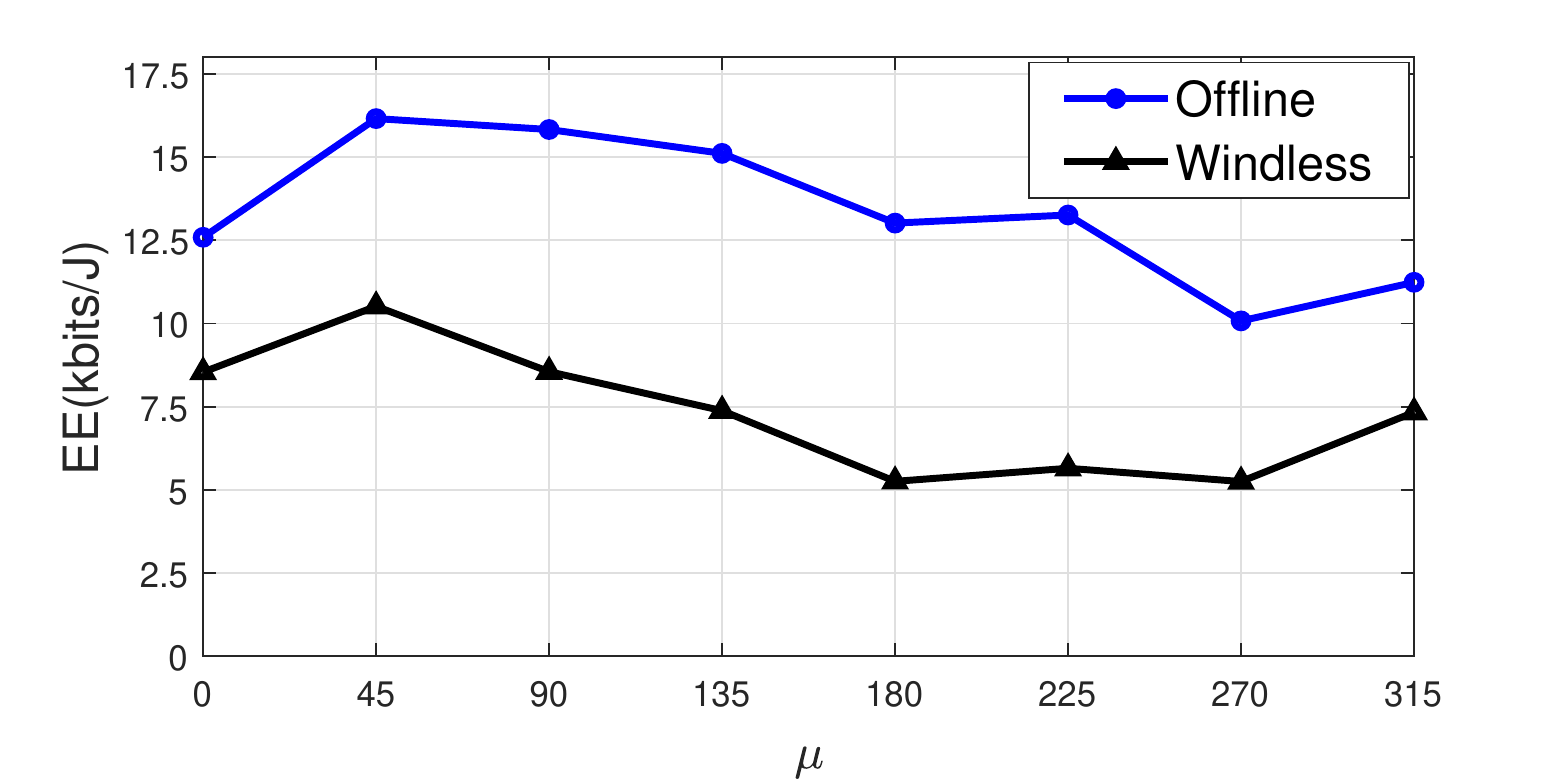}
    \caption{ EE comparison based on different $\mu$ ($\kappa = 20$, $\lambda = 10$, $c = 10$)}
    \label{fig_ee_offline_theta}
\end{minipage}
\qquad
\begin{minipage}[t]{0.47\textwidth}
    \centering
    \includegraphics[width=1\textwidth]{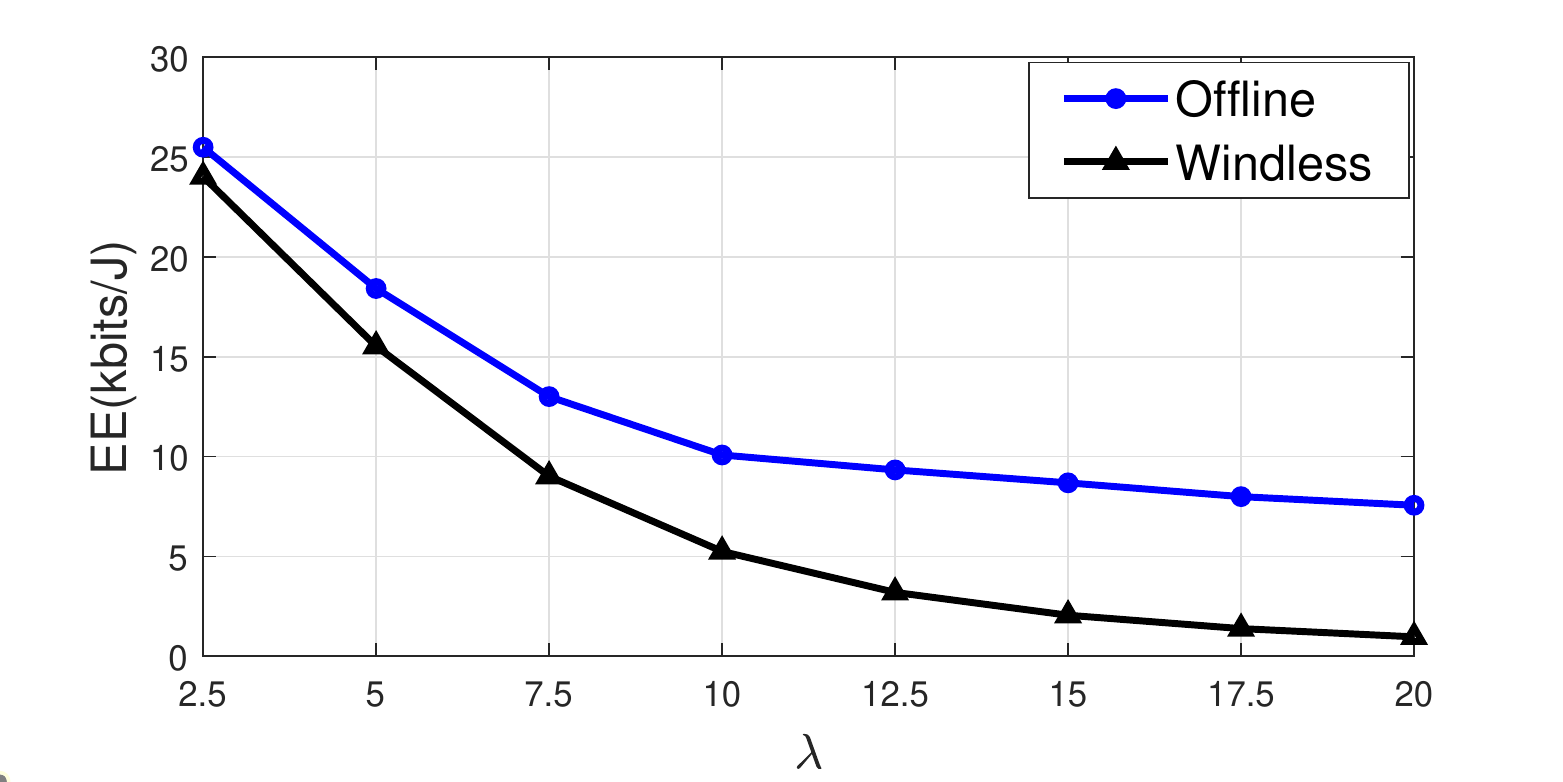}
    \caption{ EE comparison based on different $\lambda$ ($\mu = 270$, $\kappa = 20$, $c = 10$)}
    \label{fig_ee_offline_speed}
\end{minipage}
\end{figure}

 {For the proposed offline design, we use MCSAA to average the wind effect. Thus, the optimized trajectories (i.e., Fig. \ref{fig_traj_offline_theta} and \ref{fig_traj_offline_speed}) are obtained from a statistical perspective based on corresponding wind distributions. Additionally, the expected rates of the UAV are computed by the probabilistic LoS channel model as seen in \eqref{rate expectation}, and the total propulsion energy of offline design and windless scheme are computed by repeating $S = 300$ independent experiments for randomized wind and averaging their numerical results. Then, we have the EE comparisons (i.e., Fig. \ref{fig_ee_offline_theta} and \ref{fig_ee_offline_speed}) as follows.}
\par Fig. \ref{fig_traj_offline_theta} shows the optimized UAV trajectories based on different values of $\mu$. From the 3D view, we see that, the UAV tends to communicate with the GUs at a higher flight altitude in a windless environment. This is because a higher flight altitude leads to a higher LoS probability, and hence a higher EE. As approaching each GU, the UAV lowers its flight height. Because a shorter horizontal distance between the UAV and each GU leads to a higher LoS probability, and also a higher EE. To probe into the impact of wind direction, we notice the UAV trajectories from GU 2, i.e., $(500,800)$ m to GU 3, i.e., $(500,200)$ m in the 3D view. Along this direction, the UAV flight direction is approximately equal to 90 degrees as shown in the 2D view in Fig. \ref{fig_traj_offline_theta}. From the definition in Section \Rmnum{2}, $\mu$ is the expectation angle of the wind. At this point, the winds with $\mu = 90$ and $\mu = 270$ are respectively the tailwind and the headwind for the UAV. Therefore, the UAV flies lower (seen in the red trajectory) to withstand the headwind, and save more propulsion energy to achieve a higher EE. On the contrary, the UAV increases its flight height (seen in the black trajectory) to achieve a higher expected achievable rate and a higher EE with the tailwind. From the 2D view in Fig. \ref{fig_traj_offline_theta}, we see that all trajectories are slightly different horizontally. Specifically, as shown in the magnified part of the 2D view, the green trajectory is more biased in the direction of 180 degrees but the blue is opposite, due to the wind effect in the corresponding direction.

\par Fig. \ref{fig_traj_offline_speed} shows the optimized UAV trajectories based on different $\lambda$. Similar to the 3D view in Fig. \ref{fig_traj_offline_theta}, without the wind impact, the UAV tends to fly higher. Additionally, we see that the average height of the UAV decreases with a larger $\lambda$, because the UAV has to fly at a lower altitude to reduce the wind impact, which becomes stronger at a higher altitude, and save more propulsion energy. For the 2D view in Fig. \ref{fig_traj_offline_speed}, all trajectories are also slightly different similar to Fig. \ref{fig_traj_offline_theta}.

 Fig. \ref{fig_ee_offline_theta} and Fig. \ref{fig_ee_offline_speed} show the EE comparisons with varying $\mu$ and $\lambda$. As expected, under the impact of the wind, the proposed offline design based on SP greatly improves the EE compared with the windless scheme. Specifically, the EEs vary greatly over $\mu$, due to the wind impact in different directions. In addition, as $\lambda$ increases, the EEs decrease, and the offline design outperforms significantly compared with the windless scheme. Therefore, the above results demonstrate the validity of the proposed offline design.
\subsection{Performance Analysis for OBOA Design}
\begin{figure}[htbp]
    \subfloat{
        \includegraphics[width=0.48\linewidth]{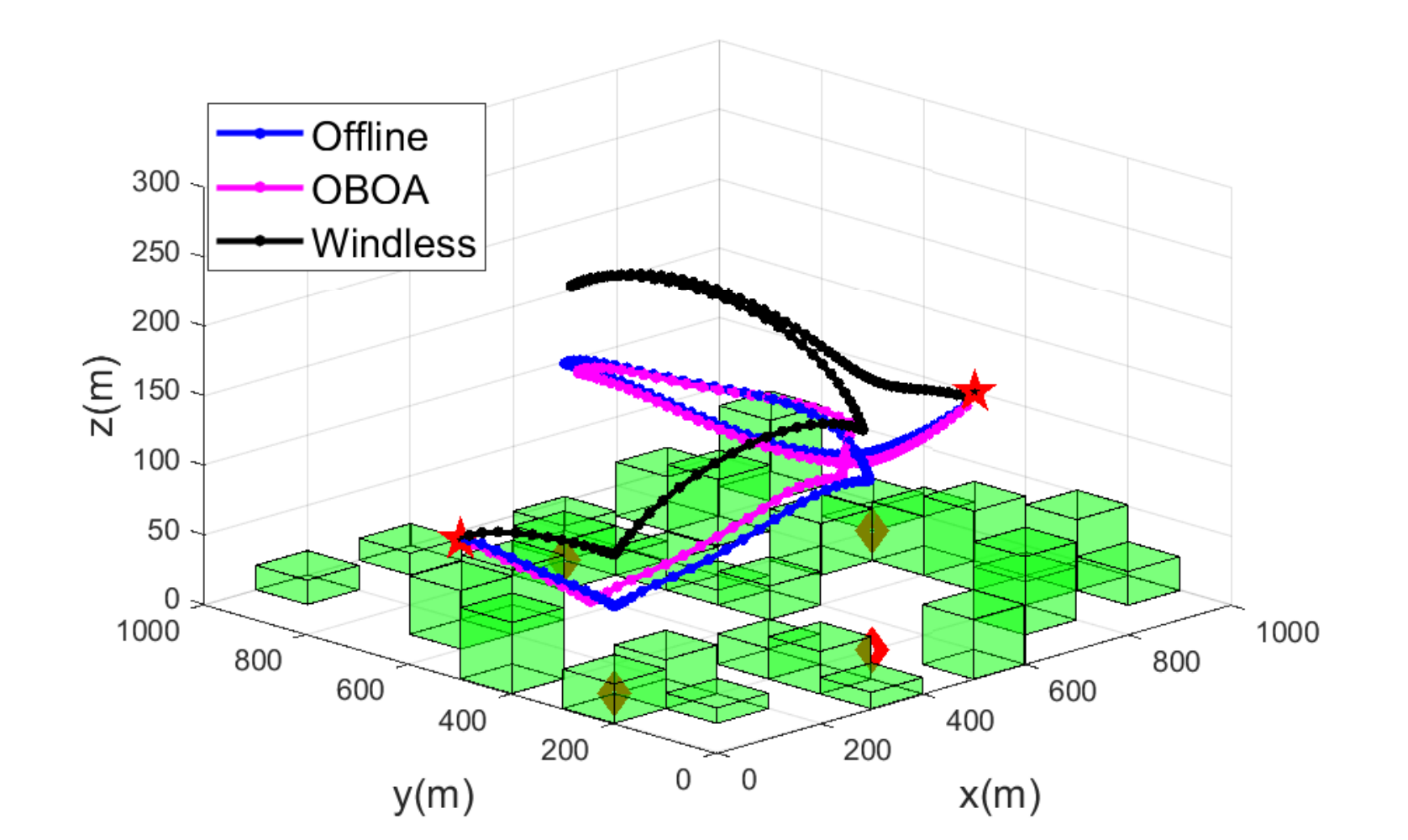}
    }
    \subfloat{
        \includegraphics[width=0.45\linewidth]{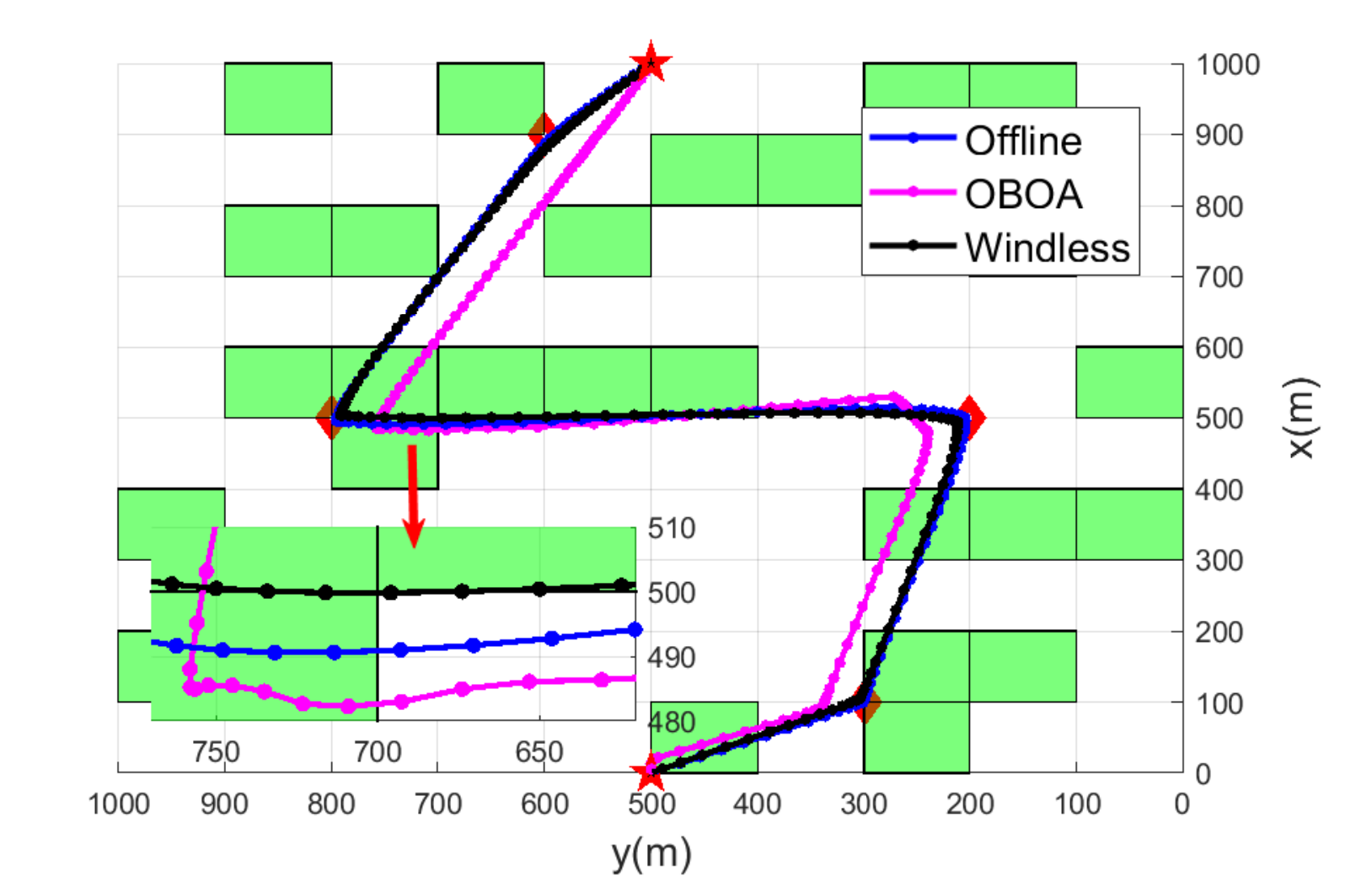}
    }
 \caption{{UAV trajectories based on different schemes ($\varepsilon_Q = 50$ m, $\mu = 180$, $\kappa = 5$, $\lambda = 10$, $c = 5$)}}
 \label{fig_traj_distance}
\end{figure}

\begin{figure}[htbp]
\centering
\subcaptionbox{$\mu = 90$\label{fig_angle_90}}
  { \includegraphics[width=0.48\textwidth]{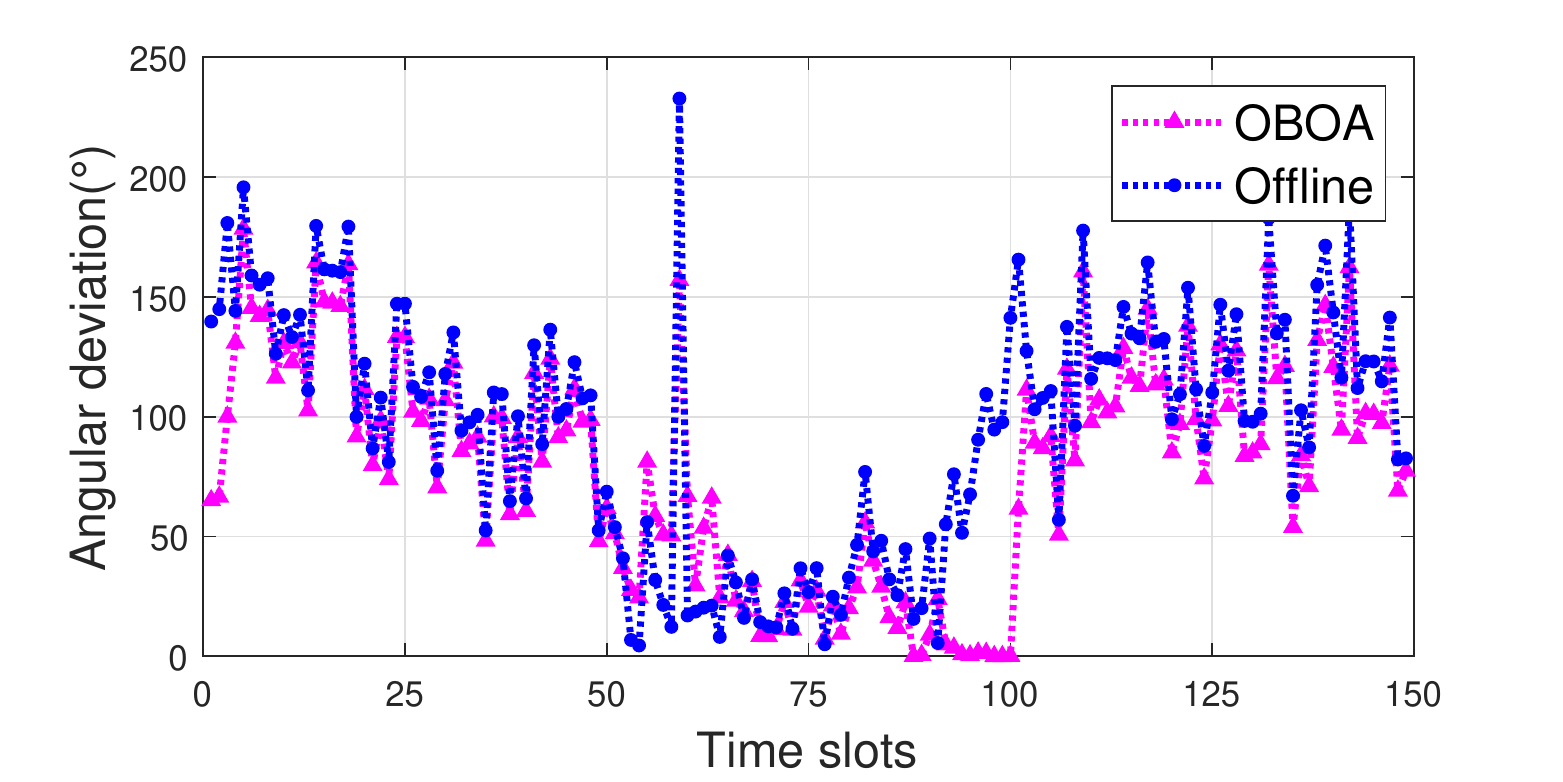}}
\subcaptionbox{$\mu = 270$\label{fig_angle_270}}
  { \includegraphics[width=0.48\textwidth]{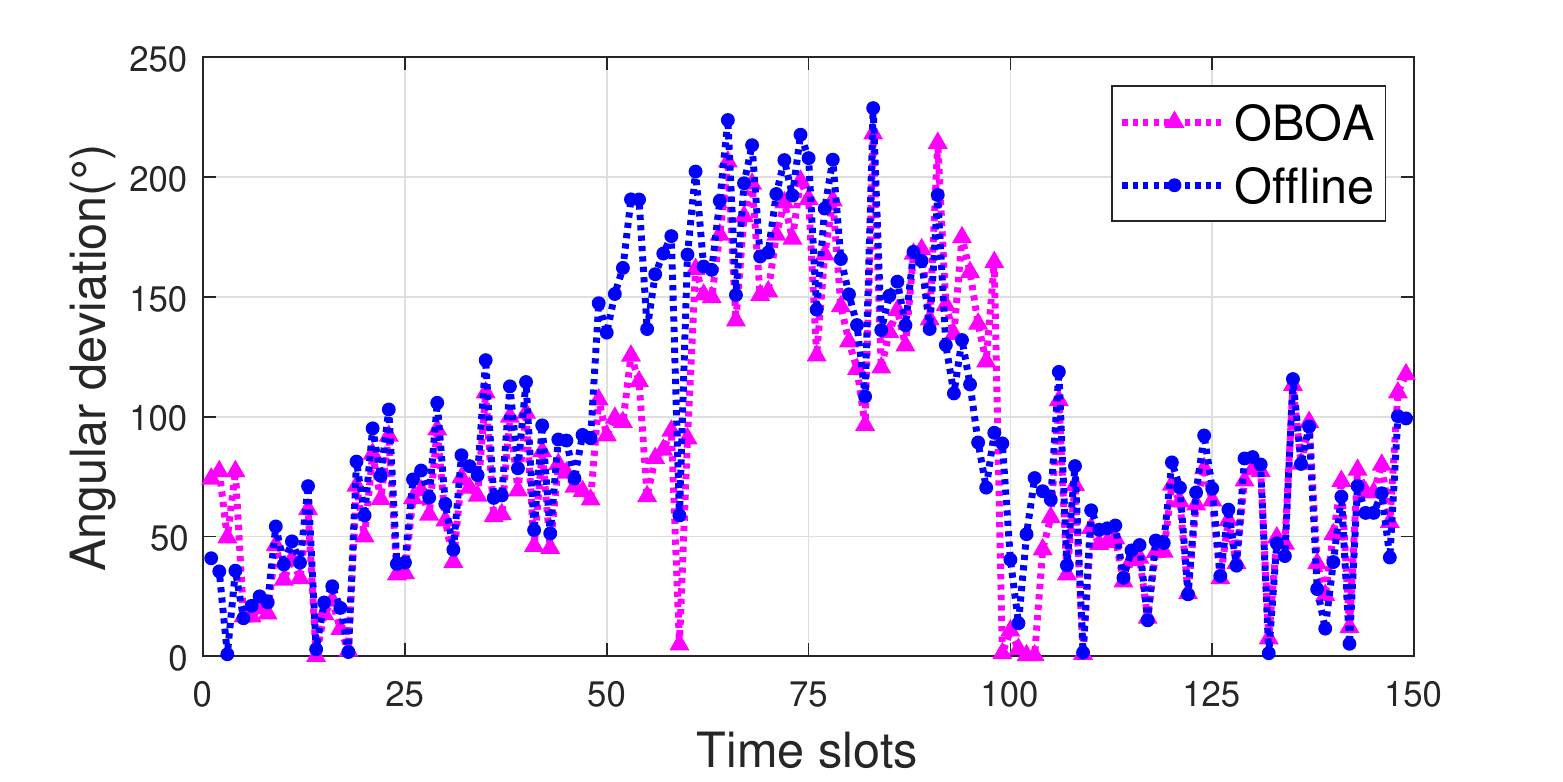}}
      \caption{ Angular deviation of the velocity with two designs w.r.t. real-time wind in each slot ($\varepsilon_Q = 100$ m, $\kappa = 5$, $\lambda = 10$, $c = 5$)}
      \label{fig_angle}
\end{figure}

\begin{figure}[htbp]
\centering
\begin{minipage}[t]{0.47\textwidth}
    \centering
    \includegraphics[width=1\textwidth]{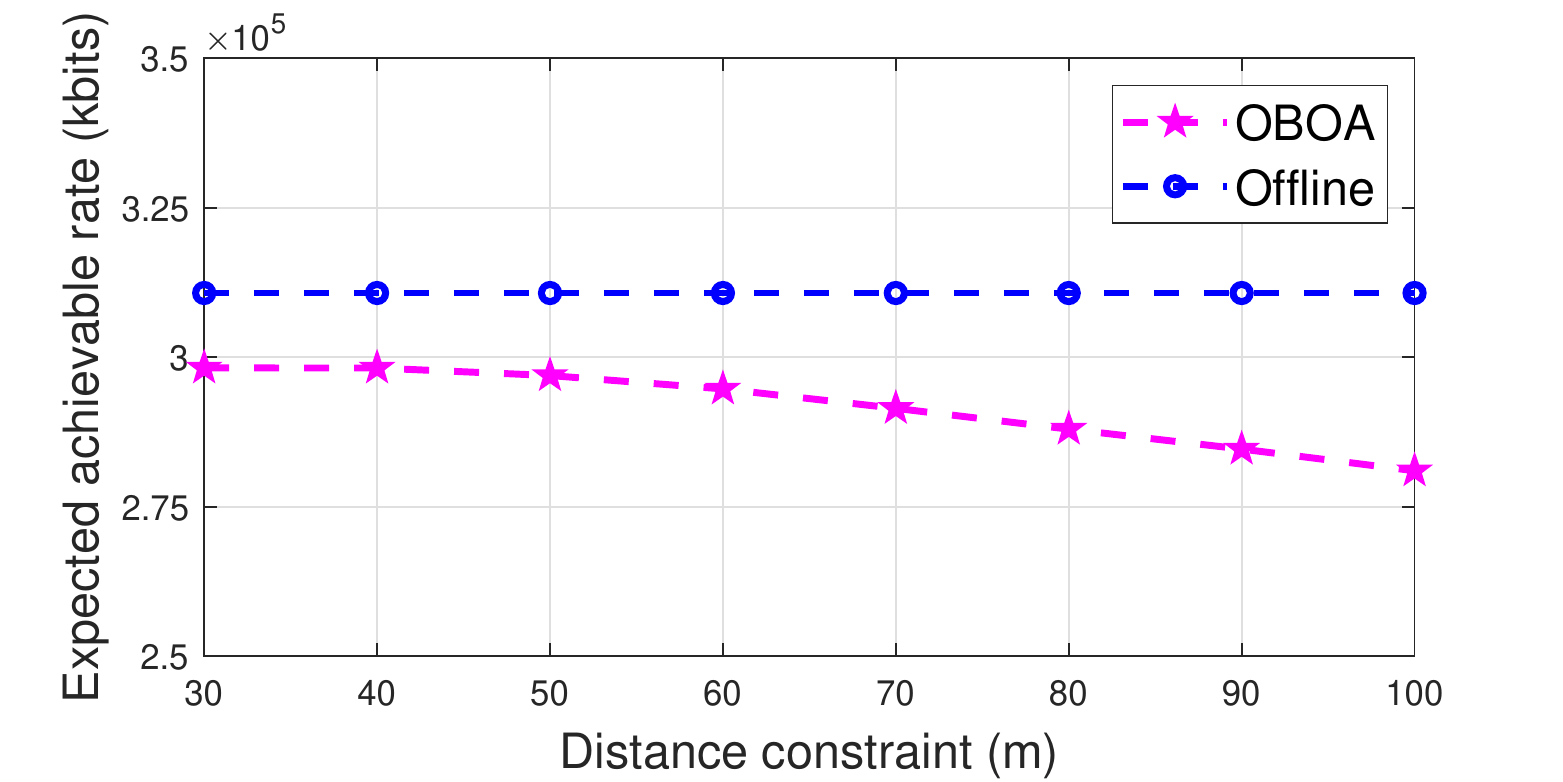}
    \caption{{ Achievable rate comparison with different $\varepsilon_Q$ ($\mu = 180$, $\kappa = 5$, $\lambda = 10$, $c = 5$)}}
    \label{fig_rate_distance}
\end{minipage}
\qquad
\begin{minipage}[t]{0.47\textwidth}
    \centering
    \includegraphics[width=1\textwidth]{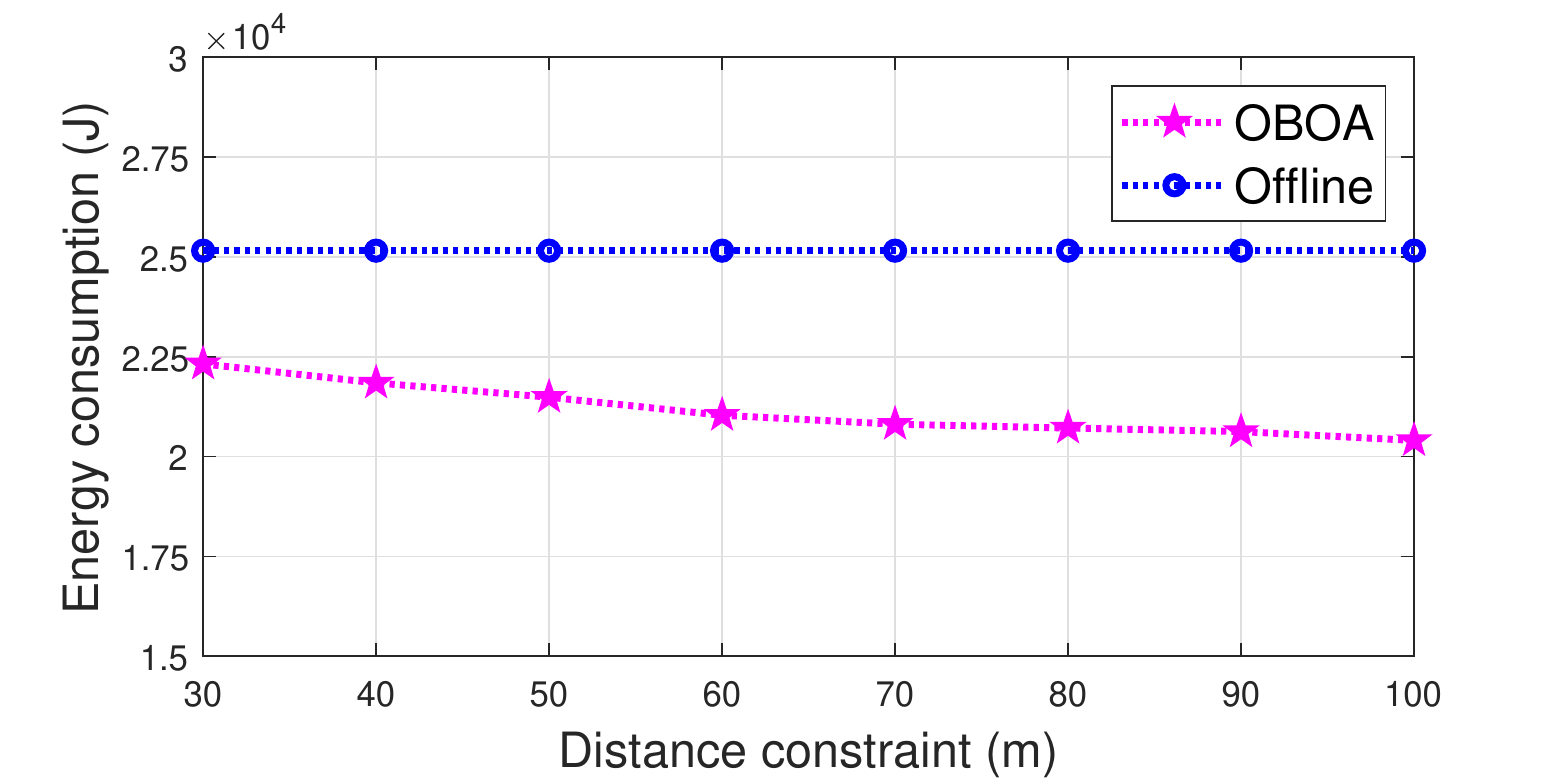}
    \caption{ Propulsion energy consumption comparison with different $\varepsilon_Q$ ($\mu = 180$, $\kappa = 5$, $\lambda = 10$, $c = 5$)}
    \label{fig_energy_distance}
\end{minipage}
\end{figure}

\begin{figure}[htbp]
\centering
\begin{minipage}[t]{0.47\textwidth}
    \centering
    \includegraphics[width=1\textwidth]{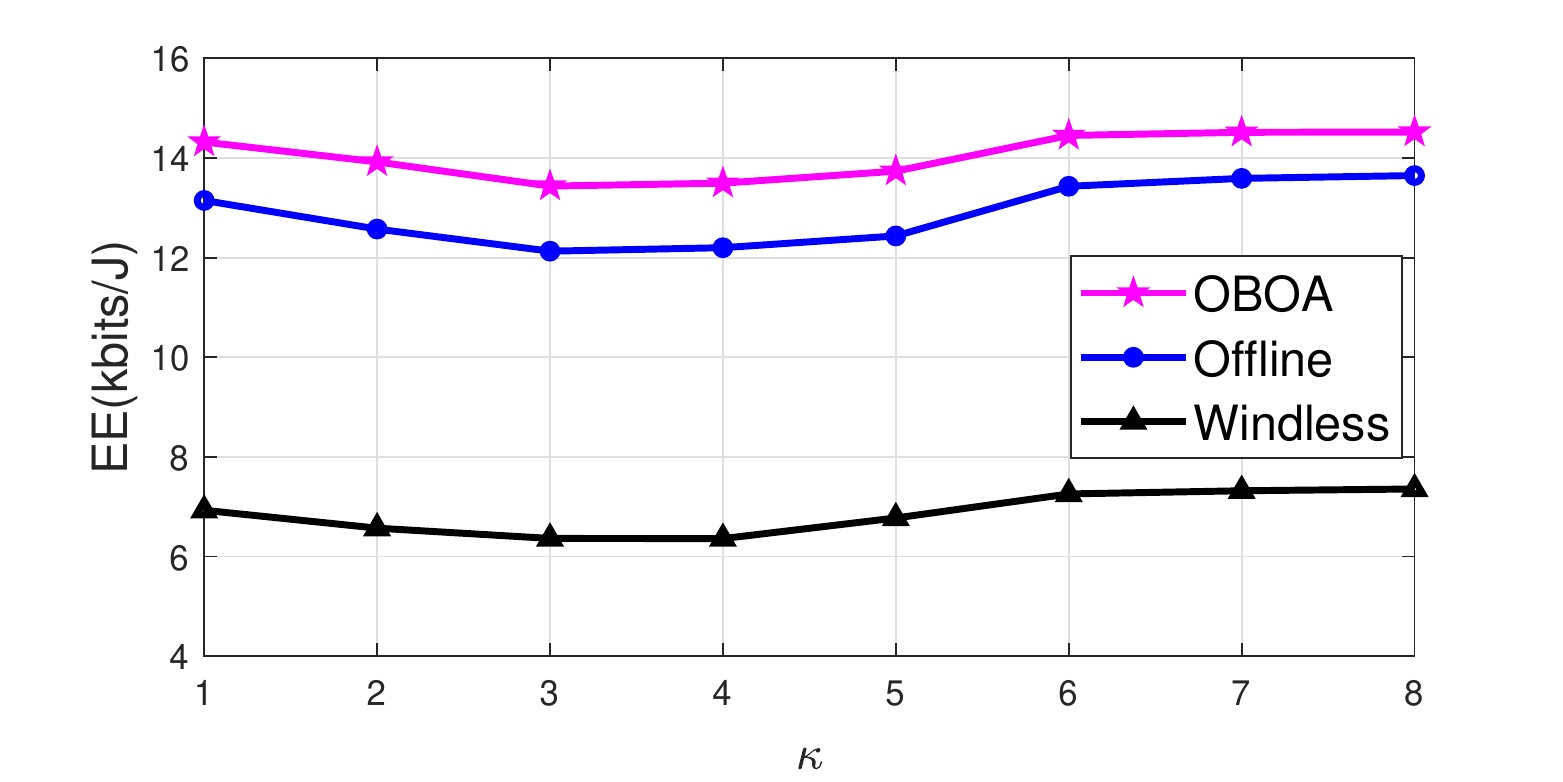}
    \caption{ {EE comparison with different $\kappa$ ($\varepsilon_Q = 100 \rm m$ , $\mu = 180$, $\lambda = 10$, $c = 5$)}}
    \label{fig_ee_kappa}
\end{minipage}
\qquad
\begin{minipage}[t]{0.47\textwidth}
    \centering
    \includegraphics[width=1\textwidth]{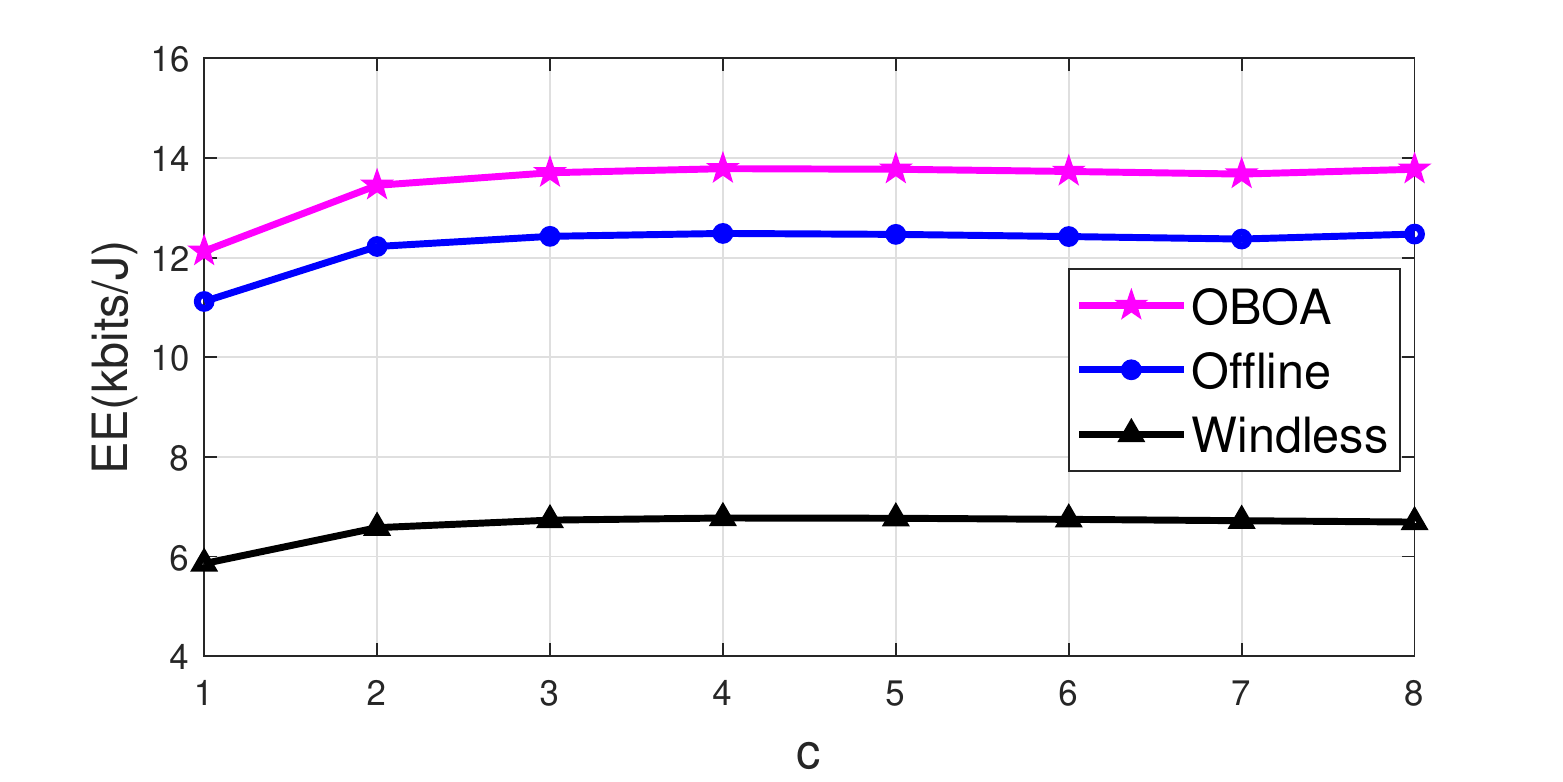}
    \caption{ {EE comparison with different $c$ ($\varepsilon_Q = 100 \rm m$ , $\mu = 180$, $\lambda = 10$, $\kappa = 5$)}}
    \label{fig_ee_shape}
\end{minipage}
\end{figure}

{For the proposed OBOA design, we consider the real environment with real-time wind effect and practical channel model. Clearly, the results of the OBOA design are highly correlated with particular parameters settings. Thus, to compare the performances of the proposed three schemes more accurately in the real environment, the achievable rates for all schemes (e.g., Fig. \ref{fig_rate_distance}) are obtained by repeating $S=300$ independent experiments for the randomized Manhattan city model and averaging their numerical results. Similarly, the total propulsion energy of all schemes (e.g., Fig. \ref{fig_energy_distance}) is computed by repeating $S = 300$ independent experiments for randomized wind and averaging their numerical results. Then, we have the EE comparisons (i.e., Fig. \ref{fig_ee_kappa} and Fig. \ref{fig_ee_shape}) as follows. Also, the optimized trajectory with the OBOA design and Manhattan city model in Fig. \ref{fig_traj_distance}, and Fig. \ref{fig_angle} are generated by an independent experiment for the randomized Manhattan city model and wind. The corresponding conclusion does not rely on any specific random parameters settings.}

\par Fig. \ref{fig_traj_distance} shows the UAV trajectories comparison. Specifically, from the 3D view, we see that the flight heights of the UAV with the OBOA and offline designs are lower than the height with the windless scheme as the analysis above. Furthermore, the optimized trajectory with the OBOA design is different from the offline design in each time slot, due to the adaptation to the real-time wind changing dramatically. To better verify this, we continue to analyze the 2D view in Fig. \ref{fig_traj_distance} and Fig. \ref{fig_angle} in the following and observe the trajectory and velocity changes of the UAV based on the online adaptation.
\par From the magnified part in the 2D view in Fig. \ref{fig_traj_distance}, we see that the optimized trajectories with the offline design and the windless scheme are both smooth, but the UAV with the OBOA design continuously adjusts its velocity in each slot. The main reason is that the trajectories with the offline design and the windless scheme are globally optimized based on MCSAA from a statistical perspective, and the wind effect on the UAV is averaged in this case. On the contrary, the trajectory with the OBOA design is constructed by a series of independent optimizations in each time slot. Furthermore, from Fig. \ref{fig_angle}, we see that the angular deviation of the OBOA design is less than that of the offline design in most time slots, yielding a smaller wind impact on UAV flight. Thus, with the OBOA design, the UAV adjusts its current flight velocity in adaptation to the real-time wind under the corresponding flight constraints, such that the OBOA design can better deal with the impact of the wind randomness, and save more propulsion energy of the UAV to achieve a higher EE.

\par Fig. \ref{fig_rate_distance} and Fig. \ref{fig_energy_distance} show the expected achievable rate and propulsion energy consumption comparisons of the UAV under varying distance constraint. Fig. \ref{fig_rate_distance} demonstrates the assumption in the online phase. The expected achievable rate with the OBOA design is lower than the rate with the offline design. However, the difference of the expected achievable rate between the OBOA and offline designs is no more than $10\%$ at best at $\varepsilon_Q = 100$ $\rm m$. The OBOA design in Fig. \ref{fig_energy_distance} saves propulsion energy of the UAV significantly compared with the offline design, and the difference of energy consumption goes up to $20\%$ at best at $\varepsilon_Q = 100$ $\rm m$. As we pay more attention to the improvement of the EE, Fig. \ref{fig_rate_distance} and Fig. \ref{fig_energy_distance} verify the effectiveness of the OBOA design.

Fig. \ref{fig_ee_kappa} and Fig. \ref{fig_ee_shape} show the EE comparisons with different levels of wind randomness by varying $\kappa$ and $c$. As is shown in Fig. \ref{fig_ee_kappa}, we see that the OBOA and offline designs outperforms the windless scheme in terms of EE, and the OBOA design is more energy-efficient than the offline design. Specifically, when $ \kappa \ge 3$, all three EEs gradually go down with decreasing $\kappa$. Because the wind varying more frequently in direction makes the UAV difficult to make timely adjustments for the flight direction. However, when $\kappa \le 3$, the EEs increase as $\kappa$ decreases. The main reason is that, a smaller $\kappa$, i.e., the direction of the wind changes more frequently, leads to a higher tailwind probability for the UAV. In this case, the UAV can save more propulsion energy compared with the condition that the direction of the wind is more concentrated. For Fig. \ref{fig_ee_shape}, the general conclusion about the EE comparison is similar to Fig. \ref{fig_ee_kappa}. {In addition, we see that the EEs go down with decreasing $c$ in Fig. \ref{fig_ee_shape}. This is because frequent changes in wind speed, especially when the wind speed becomes higher, will cause more propulsion energy consumption for the UAV.} To sum up, the above results corroborate that the proposed OBOA design is more energy-efficient, which makes the UAV better adapt to the real-time wind changing more dramatically.

\section{Conclusion}
In this paper, we investigated the effect of the stochastic wind on the UAV-enabled communication system. We first derived a 3D GPECM of a rotary-wing UAV affected by the stochastic wind. Then, we formulated a stochastic optimization problem to maximize the EE of the UAV, by jointly optimizing the 3D trajectory and user scheduling. To tackle this problem, we proposed an OBOA design yielding a statistical solution based on wind statistics and SP technique in the offline phase, and further optimized UAV instantaneous velocity in adaptation to the real-time wind in each slot based on the offline design in the online phase. Numerical results show the optimized UAV trajectories of both two phases can better adapt to the wind in changing speed and direction, and achieve a higher EE compared to the benchmark scheme. Moreover, the proposed OBOA design can be applied in the scenario with dramatic wind changes and achieves a better performance in terms of EE.
%if have a single appendix:
% or
%\appendix% for no appendix heading


{
\appendices
\section{Simplifications For the GPECM}
Since the existing models in \cite{zeng2019energy,9847346,cai2022resource} all lack the analysis of the impact of wind on the UAV propulsion power, we first simplify the GPECM into a form without considering the wind effect. Specifically, the GPECM is derived based on the 3D force analysis on a rotary-wing UAV, and winds exert impact on the UAV as shown in \eqref{D} and \eqref{T}. Hence, if we neglect the wind effect, i.e., by setting $\|\mathbf{v}_w\| = 0$ in \eqref{D} and \eqref{T}, the GPECM without considering the wind effect which has a similar form as \eqref{P_2}, can be expressed as
          \begin{align}
            \label{P_nowind}
            P = &\frac{\delta}{8}\left(\frac{\|\mathbf{T}\|}{c_T\rho A}+3\|\mathbf{v}\|^2\right)\sqrt{\frac{\rho s^2A\|\mathbf{T}\|}{c_T}} + (1+c_f)\|\mathbf{T}\|\left(\sqrt{\frac{\|\mathbf{T}\|^2}{(2\rho A)^2}+\frac{\|\mathbf{v}\|^4}{4}}-\frac{\|\mathbf{v}\|^2}{2}\right)^{\frac{1}{2}}\nonumber\\ &+m\|\mathbf{g}\|\|\mathbf{v}\|\sin{\tau_c}+\frac{1}{2}\rho S_{\rm FP}\|\mathbf{v}\|^3,
            \end{align}
where $\|\mathbf{T}\|=\|m\mathbf{a}+\frac{1}{2}\rho S_{\rm FP}\|\mathbf{v}\|\mathbf{v}-m\mathbf{g}\|$. We now clarify how \eqref{P_nowind} reduces to the existing models under different assumptions.
\subsubsection{Simplification for the GPECM to the model in \cite{zeng2019energy}}
In \cite{zeng2019energy}, a 2D propulsion energy consumption model is derived, which a function of UAV velocity and other constant fuselage parameters. Specifically, the authors of \cite{zeng2019energy} only considered the impact of the horizontal velocity and the flight drag on UAV propulsion power, and assumed that the thrust always equaled to the weight considering the horizontal flight. Thus, we have $\mathbf{T}=m\mathbf{g}$, and can omit the terms $m\mathbf{a}$ and climbing power $m\|\mathbf{g}\|\|\mathbf{v}\|\sin{\tau_c}$ in \eqref{P_nowind}. Then, based on $v_o=\sqrt{\frac{\|\mathbf{T}\|}{2\rho A}}=\sqrt{\frac{\|m\mathbf{g}\|}{2\rho A}}$ and $d_0 = \frac{S_{\rm FP}}{sA}$, \eqref{P_nowind} is recast as
         \begin{align}
         \label{P_zeng_1}
            P = &\frac{\delta}{8}\left(\frac{\|\mathbf{T}\|}{c_T\rho A}+3\|\mathbf{v}\|^2\right)\sqrt{\frac{\rho s^2A\|\mathbf{T}\|}{c_T}} + (1+c_f)\frac{\|m\mathbf{g}\|^{\frac{3}{2}}}{\sqrt{2\rho A}}\kappa\left(\sqrt{\kappa^2+\frac{\|\mathbf{v}\|^4}{4v_o^4}}-\frac{\|\mathbf{v}\|^2}{2v_o^2}\right)^{\frac{1}{2}}+\frac{1}{2}d_o\rho sA\|\mathbf{v}\|^3,
            \end{align}
         where $\kappa=\|\mathbf{T}\|/\|m\mathbf{g}\|$ denotes the thrust-to-weight ratio defined in \cite{zeng2019energy}, and we have $\kappa=1$ based the above assumption. Then, based on $\|\mathbf{T}\|=c_T\rho A v_{\rm tip}^2$ and $\hat{v} = \frac{\|\mathbf{v}\|}{v_{\rm tip}}$, \eqref{P_zeng_1} can be rewritten as 
         \begin{align}
         \label{P_zeng_2}
            P = &\frac{\delta}{8}\left(1+3\frac{\|\mathbf{v}\|^2}{v^2_{\rm tip}}\right) + (1+c_f)\frac{\|m\mathbf{g}\|^{\frac{3}{2}}}{\sqrt{2\rho A}}\left(\sqrt{1+\frac{\|\mathbf{v}\|^4}{4v_o^4}}-\frac{\|\mathbf{v}\|^2}{2v_o^2}\right)^{\frac{1}{2}}+\frac{1}{2}d_o\rho sA\|\mathbf{v}\|^3.
            \end{align}
         This is exactly the same as model (12) in \cite{zeng2019energy}.
\subsubsection{Simplification for the GPECM to the model in \cite{9847346}}
In \cite{9847346}, an improved 2D propulsion energy consumption model based on \eqref{P_zeng_2} is derived. The authors of \cite{9847346} considered the impact of thrust-to-weight ratio $\kappa$ and obtain a model w.r.t. the velocity, acceleration and direction change in the horizontal direction of the UAV. Note the assumptions of the model in \cite{9847346} are the same as in [R15] besides considering $\kappa\neq 1$. Thus \eqref{P_nowind} is reduced to 
         \begin{align}
         \label{P_dai_1}
            P = &\frac{\delta}{8}\left(1+3\frac{\|\mathbf{v}\|^2}{v^2_{\rm tip}}\right)+ (1+c_f)\frac{\|m\mathbf{g}\|^{\frac{3}{2}}}{\sqrt{2\rho A}}\kappa\left(\sqrt{\kappa^2+\frac{\|\mathbf{v}\|^4}{4v_o^4}}-\frac{\|\mathbf{v}\|^2}{2v_o^2}\right)^{\frac{1}{2}}+\frac{1}{2}d_o\rho sA\|\mathbf{v}\|^3.
            \end{align}
         From Fig. 1 in \cite{9847346}, we similarly denote $\phi$ as the angle between $\mathbf{T}$ and $Z$-axis, and denote $\varphi$ as the angle between the drag $\mathbf{D}$ and $\mathbf{F}$ (i.e., the resultant force on the UAV). Note that $\mathbf{D}$ and $\mathbf{F}$ both are forces in the horizontal direction. Then, based on $\|\mathbf{D}\|=\frac{1}{2}\rho S_{\rm FP}\|\mathbf{v}\|^3$ and the law of cosines, we take the force analysis on the UAV in both horizontal and vertical direction, yielding 
         \begin{align}
            \|\mathbf{T}\||\sin\phi| = \sqrt{\|\mathbf{F}\|^2+\|\mathbf{D}\|^2-2\|\mathbf{F}\|\|\mathbf{D}\|\cos\varphi},\quad  \|\mathbf{T}\||\cos\phi|=\|m\mathbf{g}\|.
            \end{align}
         Then, based on $|\sin\phi|^2+|\cos\phi|^2=1$ and $\kappa=\|\mathbf{T}\|/\|m\mathbf{g}\|$, we have
         \begin{align}
            \label{new_kappa}
            \kappa = \sqrt{1+\frac{4m^2\|\mathbf{a}\|^2+\rho^2 S_{FP}^2\|\mathbf{v}\|^4+4m\rho S_{FP}\mathbf{a}\mathbf{v}\|\mathbf{v}\|}{4\|m\mathbf{g}\|^2}}.
            \end{align}
         Therefore, \eqref{P_nowind} is reduced to equation (5) in \cite{9847346} by substituting \eqref{new_kappa} into \eqref{P_dai_1}.
\subsubsection{Simplification for the GPECM to the model in \cite{cai2022resource}}
In \cite{cai2022resource}, the authors proposed a 3D propulsion energy consumption model by considering the propulsion energy consumption in horizontal and vertical direction as two independent components. In addition, the authors of \cite{cai2022resource} also neglected the impact of the vertical flight drag and acceleration on the UAV propulsion power, similarly as in [R15]. Then from \cite{cai2022resource}, we similarly decompose the velocity $\mathbf{v}$ into three components in $X$, $Y$ and $Z$-axis, respectively, i.e., $\mathbf{v} = (v_x,v_y,v_z)$. Since $\tau_c$ denotes the climbing angle and $\|\mathbf{v}\|$ in \eqref{P_zeng_2} is the horizontal velocity of the UAV, we have $\|\mathbf{v}\|\sin{\tau_c}=v_z$. Thus, \eqref{P_nowind} can be recast as 
         \begin{align}
         \label{P_cai_1}
            P = &\frac{\delta}{8}\left(1+\frac{3(v_x^2+v_y^2)}{v^2_{\rm tip}}\right)+(1+c_f)\frac{\|m\mathbf{g}\|^{\frac{3}{2}}}{\sqrt{2\rho A}}\left(\sqrt{1+\frac{(v_x^2+v_y^2)^2}{4v_o^4}}-\frac{v_x^2+v_y^2}{2v_o^2}\right)^{\frac{1}{2}}\nonumber\\
            &+m\|\mathbf{g}\|v_z+\frac{1}{2}d_o\rho sA(v_x^2+v_y^2)^{\frac{3}{2}}.
            \end{align}
         Therefore, \eqref{P_nowind} is reduced to the form as equation (17) in \cite{cai2022resource}.}



\ifCLASSOPTIONcaptionsoff
  \newpage
\fi



%\bibliographystyle{IEEEtran}


% \end{thebibliography}
\bibliographystyle{IEEEtran} 
\bibliography{reference}


% \end{IEEEbiography}


%\newpage



%\vfill


\end{document}